\documentclass{aa}  

\usepackage{graphicx}
\usepackage{txfonts}
\usepackage[colorlinks=true, citecolor=blue, linkcolor=blue, urlcolor=blue]{hyperref}

\begin{document}

\title{Forecasting the power of Higher Order Weak Lensing Statistics with automatically differentiable simulations}

\author{Denise Lanzieri \inst{1}\fnmsep\thanks{ Contact: denise.lanzieri@cea.fr}
\and
Fran\c{c}ois Lanusse \inst{2}
\and
Chirag Modi \inst{3}
\and 
Benjamin Horowitz \inst{4, 5}
\and
Joachim Harnois-Déraps \inst{6}
\and
Jean-Luc Starck \inst{2}
\and
The LSST Dark Energy Science Collaboration (LSST DESC)
}
\institute{Université Paris Cité, Université Paris-Saclay, CEA, CNRS, AIM, F-91191, Gif-sur-Yvette, France
\and
Université Paris-Saclay, Université Paris Cité, CEA, CNRS, AIM, 91191, Gif-sur-Yvette, France
\and
Center for Computational Astrophysics,
Center for Computational Mathematics,
Flatiron Institute, New York, NY 10010, USA
\and
Lawrence Berkeley National Laboratory, 1 Cyclotron Road, Berkeley, 94720, CA, USA
\and
Department of Astrophysical Sciences, Princeton University, Princeton, NJ 08544, USA
\and
School of Mathematics, Statistics and Physics, Newcastle University, Herschel Building, NE1 7RU Newcastle-upon-Tyne, UK
}

\date{Received xxxx; accepted xxx}

 
  \abstract
   {}
   {We present the Differentiable Lensing Lightcone (DLL), a fully differentiable physical model designed for being used as a forward model in Bayesian inference algorithms requiring access to derivatives of lensing observables with respect to cosmological parameters. 
}
   {We extend the public FlowPM N-body code, a particle-mesh N-body solver, simulating lensing lightcones and implementing the Born approximation in the Tensorflow framework. Furthermore, DLL is aimed at achieving high accuracy with low computational costs. As such, it integrates a novel Hybrid Physical-Neural parameterisation able to compensate for the small-scale approximations resulting from particle-mesh schemes for cosmological N-body simulations.
We validate our simulations in an LSST setting against high-resolution $\kappa$TNG simulations by comparing both the lensing angular power spectrum and multiscale peak counts. We demonstrate an ability to recover lensing $C_\ell$ up to a 10\% accuracy at $\ell=1000$ for sources at redshift 1, with as few as $\sim 0.6$ particles per Mpc/h. 
As a first use case, we use this tool to investigate the relative constraining power of the angular power spectrum and peak counts statistic in an LSST setting.
Such comparisons are typically very costly as they require a large number of simulations, and do not scale well with the increasing number of cosmological parameters.  
As opposed to forecasts based on finite differences, these statistics can be analytically differentiated with respect to cosmology, or any systematics included in the simulations at the same computational cost of the forward simulation. }
   {We find that the peak counts outperform the power spectrum on the cold dark matter parameter $\Omega_c$, on the amplitude of density fluctuations $\sigma_8$, and on the
amplitude of the intrinsic alignment signal $A_{IA}$.}
   {}


\maketitle
%

\section{Introduction}
Weak gravitational lensing by Large Scale Structures (LSS) is one of the key probes to test cosmological models and gain insight into constituents of the Universe. The upcoming stage-IV surveys, such as the Legacy Survey of Space and Time (LSST) of the Vera C. Rubin Observatory \cite{ivezic2019lsst}, the Nancy Grace Roman Space Telescope \citep{spergel2015wide}, and  the Euclid Mission \citep{laureijs2011euclid}, will provide measurements of billions of galaxy shapes with unprecedented accuracy, which in turn will lead to tight constraints on dark energy models \citep[e.g.][]{mandelbaum2018lsst}.

With the increased statistical power of these surveys comes the question of their optimal analysis. Traditional cosmological analysis rely on measurements of the two-point statistics, either the shear two-point correlation functions or its Fourier transform, the lensing power spectrum. However, the two-point statistics are only optimal for Gaussian fields, and do not fully capture the non-Gaussian information imprinted in the lensing signal at the scales that future surveys will be able to access (e.g. information encoded in the peaks and in the filamentary features of the matter distribution).

 This has led to the introduction of a number of  higher-order statistics to access the non-Gaussian information from weak lensing data: 
the weak lensing one point PDF \citep{liu2019constraining, uhlemann2020fisher, boyle2021nuw}, lensing peak counts \citep{liu2015cosmology, liu2015cosmological, lin2015new, kacprzak2016cosmology, peel2017cosmological, shan2018kids, martinet2018kids, ajani2020constraining, harnois2021cosmic, zurcher2022dark}, Minkowski functionals \citep{kratochvil2012probing, petri2013cosmology}, moments of mass maps \citep{gatti2021dark}, wavelet and scattering transforms \citep{ajani2021starlet, cheng2021weak}, and 3 point statistics \citep{takada2004cosmological, semboloni2011weak, rizzato2019tomographic, halder2021integrated}.

Recently, machine learning-based methods that broadly fall in the category of Simulation-Based Inference (SBI) \citep{fluri2019cosmological, kacprzak2022deeplss, fluri2021cosmological, jeffrey2021likelihood, fluri2022full}, and Bayesian forward-modeling frameworks \citep{porqueres2021bayesian, sarma2022map} have also been introduced to attempt to fully account for the non-Gaussian content in the weak lensing signal. Unlike the methods described above, these approaches are designed to access the full field-level information content. Even though these techniques are asymptotically theoretically optimal in terms of information recovery, they still suffer from significant limitations.

SBI methods are characterized by the absence of an analytical model to describe the observed signal and instead rely on learning a likelihood from simulations. Modern approaches employ deep learning-based density estimation methods to model the likelihood without the need to make any Gaussianity assumptions, and as such have 
drawn the attention of the community \citep{alsing2018massive, jeffrey2021likelihood}.  

A key element common to most of these methodologies is their ability to benefit from gradient information. 
For example, \cite{porqueres2021bayesian} present a Bayesian hierarchical approach to infer the cosmic matter density field simultaneously with cosmological parameters using the Hamiltonian Monte Carlo (HMC) algorithm to explore the full high-dimensional parameter space. The HMC algorithm exploits the information encoded in the gradients of the joint likelihood function to inform each update step and hence requires the derivatives of the forward model. 

 As a different class of examples, \cite{makinen2021lossless} demonstrate that the full information content of a cosmological field can be represented by optimal summaries, allowing for likelihood-free and near-exact posteriors for cosmological parameters. In particular, they use neural networks trained on simulations to maximise the Fisher information, which requires having access to derivatives of the simulation model to compute a Fisher matrix.  

One possible way of evaluating these gradients with respect to cosmological parameters is by using numerical differentiation, e.g. computing a finite difference of a given statistic by varying the simulation fiducial values by a small amount.
However, numerical differentiation is expensive in terms of computational resources and simulation time, and also requires hyperparameter tuning for the step size used in the finite difference scheme. An alternative option consists in computing the gradient analytically. Both these solutions are faced with limitations: the first approach gets computationally intractable in high dimensions, while the analytic gradients are sometimes impossible to estimate.
All of these new techniques for cosmological analyses make the development of fast and differentiable simulations necessary.

\cite{bohm2021madlens} developed MADLens, a CPU-based python package for producing non-Gaussian lensing convergence maps. These maps are differentiable with respect to the initial conditions of the underlying numerical simulations and to the cosmological parameters $\Omega_m$ and $\sigma_8$. In this paper,  we aim to efficiently compute gradients that benefit the development of new inference algorithms for weak lensing surveys. To do this we extend the framework of the FlowPM package \cite{modi2021flowpm} by implementing the Born approximation and simulating lensing lightcones in the Tensorflow framework. TensorFlow is a tensor library that includes the ability to perform automatic differentiation. Automatic differentiation enables us to compute gradients exactly as opposed to finite differences, which only provide approximate gradients. Specifically, TensorFlow implements the backpropagation algorithm to compute gradients, i.e. first it creates a graph (e.g. data structures representing units of computation), then it works backward through the graph by applying the chain rule at each node. Unlike MADlens, our tool is GPU-based and provides derivatives with respect to all the cosmological parameters. There are also differences in the implementation of various functions to improve accuracy and speed.

We validate our simulations against the cosmological N-body simulations $\kappa$TNG \citep{osato2021kappatng} by comparing both the lensing angular power spectrum and multiscale peak counts.

In particular, as a first application, we show how the differentiability of numerical simulations can be exploited to evaluate the Fisher Matrix. Then, we compare the constraining power of two map-based weak lensing statistics: the lensing power spectrum and peak counts and investigate the degeneracy in high dimensional cosmological and nuisance parameter space through Fisher forecasts. 

This paper is structured as follows: in \ref{weak_lesning_modeling} we briefly review the weak lensing modeling including the theoretical framework and the summary statistics used in this work. In \autoref{Fast_and_Differentiable_Lensing_Simulations} we introduce the numerical simulations illustrating the numerical methods used to generate mock WL maps. In \autoref{Validating_simulations_for_LSST_Y} we validate the simulations by comparing the statistics from our simulations and $\kappa$TNG-Dark ones.
The Fisher forecast formalism and the survey and noise setting are shown in \autoref{Application_Fisher_forecast_for_HOS}. We finally discuss our results and present our conclusions afterward, in \autoref{Discussion} and \autoref{Conclusions}.


\section{Weak Lensing Modeling}\label{weak_lesning_modeling}
\subsection{{Cosmic shear}}
Weak gravitational lensing is a powerful probe to infer the distribution of matter density between an observer and a source.
The effect of gravitational lensing 
can be quantified in term of the separation vector \textbf{x} between two light rays separated by an angle $\boldsymbol{\theta}$:
\begin{align}\label{sep}
   & \textbf{x}(\boldsymbol{\theta},\chi) =
    f_k(\chi)\boldsymbol{\theta} + \\
   - \frac{2}{c^2}
   \int_0^{\chi}
    &  \text{d}\chi'
    f_k(\chi-\chi')
    [ \boldsymbol{\nabla_{\bot}}\Phi( \textbf{x}(\boldsymbol{\theta},\chi'),\chi')-
    \boldsymbol{\nabla_{\bot}}\Phi^{(0)}(\chi')
    ],
\end{align}
with $\Phi$ and $\Phi^{0}$ the gravitational potential along the two light rays, $f_k(\chi)$ and $\chi$ the angular and radial comoving distance.

Formally, the effect of the linearized lens mapping is described by the Jacobian matrix:
\begin{equation}
\mathcal{A}(\boldsymbol{\theta},\chi)=
\frac{1}{f_{k}(\chi)}
\frac{\partial \textbf{x}}{\partial \boldsymbol{\theta}}.
\end{equation}
In the limit of weak-field metric (small $\Phi$), the integral in \autoref{sep} can be approximated by considering the series expansion in power of $\Phi$ and truncating the series at the first term.
With these assumptions, given that $\nabla_{\bot}\boldsymbol{\Phi}^0$ is not dependent from $\boldsymbol{\theta}$, the Jacobian matrix can be written as:
\begin{equation}
    \mathcal{A}_{ij}(\boldsymbol{\theta},\chi)
     =\delta_{ij}-\frac{2}{c^2}
    \int_0^{\chi} d\chi'
     \frac{f_k(\chi-\chi')f_k(\chi')}{f_k(\chi)}
     \Phi_{ij}(f_k(\chi')\boldsymbol{\theta},\chi').
\end{equation}
This, also known as Born approximation, corresponds to integrating the potential gradient along the unperturbed ray.
If we define the 2D potential, the \textit{lensing potential} as:
\begin{equation}
    \psi(\boldsymbol{\theta},\chi) \equiv
 -\frac{2}{c^2}
 \int_0^{\chi} d\chi'
   \frac{f_k(\chi-\chi')f_k(\chi')}{f_k(\chi)f_k(\chi')}
  \Phi(f_k(\chi')\boldsymbol{\theta},\chi')
\end{equation}
 the Jacobi matrix can be written as:
\begin{equation}
    \mathcal{A}_{ij}=\delta_{ij}-\partial_i \partial_j\psi.
\end{equation}

From the parametrization of the symmetrical matrix $\mathcal{A}$, we can define the spin-two shear $\gamma=(\gamma_1,\gamma_2)$ and the scalar convergence field, $\kappa$. 
Hence, the convergence and the shear are defined as the second derivative of the potential:
\begin{equation}\label{kshear}
    \kappa=\frac{1}{2}(\partial_1\partial_1+\partial_2\partial_2)\psi;  
\end{equation}
\begin{equation}
 \gamma_1=\frac{1}{2}(\partial_1\partial_1-\partial_2\partial_2)\psi; \  \  \
 \gamma_2=\partial_1\partial_2\psi;   
\end{equation}
the two fields $\gamma$ and $\kappa$ describe the distortion in the shape of the image, and the change in the angular size, respectively.
By combining the 2D Poisson equation with the \autoref{kshear}, the convergence $\kappa$ becomes:
\begin{equation}\label{born_approx}
    \kappa_{born}(\boldsymbol{\theta})= \frac{3H_0^2 \Omega_m}{2c^2}
    \int_0^{\chi_s} 
    \frac{d\chi}{a(\chi)}
    g(\chi)
    \delta(f_k(\chi)\boldsymbol{\theta},\chi),
\end{equation}
where we define the \textit{lensing efficiency}:
\begin{equation}
   g(\chi) \equiv
  \int_{\chi}^{\chi_{lim}} \text{d}\chi'
   n(\chi')
    \frac{f_k(\chi'-\chi)}{f_k(\chi')}.
\end{equation}

Thus, the Born–approximated convergence can be interpreted as the integrated total matter density along the line of sight, weighted by the distance ratios and the normalised source galaxy  distribution $n(\chi)$d$\chi=n(z)$d$z$.

\subsubsection{Intrinsic alignments (NLA)}

The galaxy ellipticity observed by a telescope can be decomposed in the cosmic shear signal $\gamma$ and the intrinsic ellipticity of the source $\epsilon^{int}$, where the latter is the combination of the alignment term $\epsilon^{IA}$ and the random component $\epsilon^{ran}$.

Different theoretical models have been proposed to describe the physics of Intrinsic Alignments (IA), such as the Non-Linear tidal Alignment model (NLA) (e.g. in \citet{Bridle_2007}), the tidal torquing model \citep{PhysRevD.70.063526, 10.1046/j.1365-8711.2001.04105.x}, or the combination of both the Tidal Alignment and Tidal Torquing model (TATT)  \citep{PhysRevD.100.103506}.
We model the IA effect using the NLA description \citep{harnois2021cosmic}, i.e. assuming a linear coupling between the intrinsic galaxy shapes and the non-linear projected tidal fields $s_{ij}$:
\begin{equation}\label{complex_e}
    \epsilon_{1}^{IA}=- \frac{A_{IA}\bar{C}_1\bar{\rho}(z=0)}{D(z)} (s_{xx}-s_{yy}), \ \ 
      \epsilon_{2}^{IA}=-\frac{2A_{IA}\bar{C}_1\bar{\rho}(z=0)}{D(z)} s_{xy},
\end{equation}
from which the observed ellipticities are computed as:
 \begin{equation}
    \boldsymbol{\epsilon}^{obs}=
    \frac{\boldsymbol{\epsilon}^{int}+\textbf{g}}{1+\boldsymbol{\epsilon}^{int,*}\textbf{g}}, \ \  \text{with}  \ \
    \boldsymbol{\epsilon}^{int}=
    \frac{\boldsymbol{\epsilon}^{
     IA}+\boldsymbol{\epsilon^{ran}}}
     {1+\boldsymbol{\epsilon}^{IA,*}\boldsymbol{\epsilon^{ran}}}.
\end{equation}
The $A_{IA}$ term in \autoref{complex_e} defines the strength of the tidal coupling, $\bar{C}_1$ is a constant
calibrated in \citet{brown2002measurement}, $D(z)$ is the linear growth function and $\bar{\rho}$ is the matter density.
 
\subsection{Lensing Summary Statistics }\label{stat}
To extract the cosmological information from the simulated $\kappa-$maps, we use two summary statistics for weak lensing observable: the angular power spectrum and the starlet peak counts \citep{lin2016new}. 

\subsubsection{Angular Cls
}
Second-order statistics, both in the form of shear 2-point correlation function $\xi_{\pm}(\theta)$, or its counterpart in Fourier space, the angular power spectrum $C_{\ell}$, have been widely used to extract the cosmological information from weak lensing surveys. \
In the Limber approximation, the angular power spectrum of the convergence field for a given tomographic bin can be computed as:
\begin{equation}
    C_{\kappa}(\ell)=\frac{9\Omega_m^2H_0^4}{4c^4}
    \int_0^{\chi_{lim}} \text{d}\chi 
    \frac{g^2(\chi)}{a^2(\chi)}P_{\delta}
    	\left( k=\frac{\ell}{f_K(\chi)},\chi \right),
\end{equation}
where $P_{\delta}$ defines the matter power spectrum of the density contrast.

The Intrinsic Alignment (IA) signal adds an excess correlation to the two-point shear correlation function (also known as cosmic shear GG or shear-shear correlation) with two terms: 1) the intrinsic-intrinsic (II) term, tracing the correlation of the intrinsic shape of two galaxies  and 2) and the intrinsic-shear coupling (GI) term, describing the correlation between the intrinsic ellipticity of one galaxy with the shear of another galaxy \citep{kilbinger2015cosmology}.
The matter power spectra for the IA terms are defined as:
\begin{align}
    P_{II}(k,z)=& \left ( \frac{A_{IA}\bar{C}_1\bar{\rho}(z)}{\bar{D}(z)} \right)^2 a^{4}(z)P_{\delta}(k,z) 
    \\
    P_{GI}(k,z)=&  \frac{A_{IA}\bar{C}_1\bar{\rho}(z)}{\bar{D}(z)} a^{2}(z)P_{\delta}(k,z),
\end{align}
where $\bar{D}(z) \equiv D(1+z)$ \citep{harnois2021cosmic}.
Under the Limber
approximation the projected angular power spectra for the IA terms become:
\begin{align}
    C_{II}=& \int_0^{\chi_{lim}} \text{d}\chi 
    \frac{n^2(\chi)}{a^2(\chi)}P_{II}(k,\chi), \\
    C_{GI}= \frac{3\Omega_m H_0^2}{2c^2} & \int_0^{\chi_{lim}} \text{d}\chi 
    \frac{g(\chi)n(\chi)}{a(\chi)}P_{GI}(k,\chi).
\end{align}

\subsubsection{Wavelet peak counts}
\paragraph{Wavelet Transform} The wavelet transform has been widely used in analysing astronomical images due to its ability to decompose astronomical data into components at different scales. 
This multiscale approach is well-suited for the study of astronomical data since their complex hierarchical structure. A wavelet function $\psi(x)$ is a function that satisfies the admissibility condition:
\begin{equation}
    \int_{\mathbb{R}^{+}} |\hat{\psi}(k)|^2 \frac{dk}{k}<\infty,
\end{equation}
where we indicate with $\hat{\psi}(k)$ the Fourier transform of $\psi(x)$, with $\int\psi(x)dx=0$ in order to satisfy the admissibility condition. 
A given signal is decomposed in a family of scaled and translated functions:
\begin{equation}
    \psi_{a,b}(x)=\frac{1}{\sqrt{a}}\psi \left( \frac{x-b}{a}\right),
\end{equation}

where $\psi_{a,b}$ are the so-called \textit{daughter wavelets}, scaled and translated version of the \textit{mother} wavelet, with $a$ and $b$ scaling and translation parameters.
The continuous wavelets transform is defined from the projections of a function $f \in L_2(\mathbb{R})$ onto the family of daughter wavelets. The coefficients of this projection represent the wavelet coefficient, obtained by :
\begin{equation}
    W_f(a,b)=\int_{\mathbb{R}} f(x)\psi^{*}_{a,b}(x)dx=\frac{1}{\sqrt{a}}\int_{\mathbb{R}}f(x)\psi^{*}\left( \frac{x-b}{a}\right)dx ,
\end{equation}
with $\psi^{*}$ the complex conjugate of $\psi$, and $ \forall a \in \mathbb{R}^{+}$, $ b \in \mathbb{R.}$  
In this work, we filter the original convergence maps with the starlet transform, an isotropic and undecimated (i.e. not down-sampled)  wavelet transform, suited for astronomical applications where objects are mostly more or less isotropic \citep{4060954}.

It decomposes an image $c_0$ as the sum of all the wavelet scales and the coarse resolution image $c_J$:
\begin{equation}\label{wav_des}
    c_0(x,y)=c_J(x,y)+\sum_{j=1}^{J_{max}} w_j(x,y)
\end{equation}
where $J_{max}$ is the maximum number of scales and $w_j$ is the wavelet images showing the details of the original image at dyadic scales with a spatial size of $2^j$ pixels and $j = J_{max} + 1$. 

The starlet wavelet function is a specific translational invariant wavelet transform:
\begin{equation}
   \frac{1}{4} \Psi \left ( \frac{x_1}{2}, \frac{x_2}{2} \right )=
    \phi (x_1,x_2)- \frac{1}{4} \phi  \left ( \frac{x_1}{2},  \frac{x_2}{2} \right )
\end{equation}
specified by an isotropic scaling function $\phi$, that,  for astronomical application, can be defined as a B-spline of order 3:
\begin{equation}
\phi_{1D}(x)= \frac{1}{12} (|x-2|^3-4|x-1|^3+6|x|^3-4|x+1|^3+|x+2|^3 ).
\end{equation}

The N-dimensional scaling functions can be built starting from the separable product of N $\phi_{1D}$: $\phi(x_1,x_2)=\phi_{1D}(x_1)\phi_{1D}(x_2)$. 
Each set of wavelet coefficients $w_j$ is obtained as the convolution of the input map with the corresponding wavelet kernel. For a full description of the starlet transform function, see \citet{4060954} and \citet{10.5555/1830428}.

\paragraph{Peak counts}It has been shown that it is necessary to go beyond second-order statistics to fully capture the non-Gaussian information encoded in the peaks of the matter distribution 
\citep{1997A&A...322....1B, 1997ApJ...484..560J,1999A&A...342...15V,2003A&A...397..809S}. Several studies have shown that the weak-lensing peak counts provide a way to capture information from non-linear structures that is complementary to the information extracted by power spectrum \citep{lin2015new, peel2017cosmological, ajani2020constraining, harnois2021cosmic, zurcher2022dark}.
The peaks identify regions of weak lensing map where the density value is higher, in this way they are particularly sensitive to massive structures.
There are two different ways to record weak lensing peaks: as 1) local maxima of the signal-to-noise field or 2) local maxima of the convergence field. In both cases, they are defined as pixels of larger value than their eight neighbors in the image.

\section{Fast and Differentiable Lensing Simulations}\label{Fast_and_Differentiable_Lensing_Simulations}
Analytical models with which to predict the observed signals do not exist for most higher-order summary statistics. 
To circumvent this issue, one approach is to rely on generating a suite of numerical simulations.
In the following section, we introduce our weak lensing map simulation procedure, including a description of the N-body simulator and the lightcone construction.

\subsection{Differentiable Particle-Mesh N-body simulations}
\subsubsection{FastPM/FlowPM}

Numerical simulations provide a practical way to model the highly nonlinear universe and extract cosmological information from observation at different scales. 

However, collision-less N-body simulations typically require significant computational effort in terms of time and CPU/GPU power, in particular, computing the gravitational interactions between the particles is typically the most time-consuming aspect and where most of the approximations are done. 
For that reason, several quasi N-body schemes have been developed to reduce the simulation time and the computational cost of full numerical simulations. 
Our weak lensing simulation tool is mainly based on the FastPM algorithm \citep{2019ascl.soft05010F} and its FlowPM \citep{modi2021flowpm} implementation which provides a fast Particle-Mesh (PM) solver estimating the gravitational forces by computing Fast Fourier Transforms on a 3D grid.

\subsection{Automatic differentiation through black-box ODE solvers}\label{Backpropagation_of_ODE_solutions}
In this work, we extend the FlowPM approach by implementing the time integration of the Ordinary Differential Equations
(ODEs) that describe the gravitational evolution of the particles in the simulation using a black-box ODE integrator. This is in contrast to the leapfrog integration method used in FastPM. One reason for this change is that adaptive ODE solvers are able to automatically adjust the time step of the simulation based on the desired accuracy for the result. Another reason for this approach is that modern automatic differentiation frameworks like TensorFlow provide automatically differentiable solvers which significantly reduce the memory footprint of the simulation when computing the gradients, as will be described below.

We begin by describing the set of equations used in the simulation:
\begin{equation}
    \left\{ \begin{array}{ll}
        \frac{d \mathbf{x}}{d a} & = \frac{1}{a^3 E(a)} \mathbf{v} \\
        \frac{d \mathbf{v}}{d a} & =  \frac{1}{a^2 E(a)} F(\mathbf{x}, a), \\
    \end{array} \right.
\end{equation}
with $\mathbf{x}$ and $\mathbf{v}$ the position and the velocity of the particles, \textit{a} the cosmological scale factor, $E(a)$ the ratio between the Hubble expansion rate and the Hubble parameter and $\mathbf{F}$ the gravitational force experienced by the dark matter particles in the mesh.

To evaluate the gradients of the solution with respect to input cosmological parameters, it is therefore necessary to  back-propagate through the ODE solver. Very recently the adjoint sensitivity method \cite{chen2018neural, pontryaginmathematical} has gained a lot of attention in the field of deep learning, as it allows to compute these gradients by solving a second ODE backward in time and treat the ODE solver as a black box. 
 
Consider an $\text{ODESolve}(\textbf{z}(t_0),f,t_0,t_1, \theta)$, 
 where $z(t)$ is the state variable, $f$ the function modeling the dynamics, $t_0$ the start time, $t_1$ the stop time and $\theta$ the dynamic parameter. The function  $\mathcal{F}$ of its output
 \begin{equation}
     \mathcal{F}(\textbf{z}(t_1))=   \mathcal{F} (\text{ODESolve}(\textbf{z}(t_0),f,t_0,t_1, \theta))
 \end{equation}
can be differentiated with respect to the input $\theta$.
 First, we need to compute the \textit{adjoint} $\textbf{a}(t)=\partial \mathcal{F}/ \partial \textbf{z}(t)$, i.e. the gradient of $\mathcal{F}$ respect to the hidden state $\textbf{z}(t)$.  
 Then, we can determine the dynamics of the adjoint through:
 \begin{equation}
     \frac{d \textbf{a}(t)}{dt}=-\textbf{a}(t)^{ \mathsf{T}}\frac{\partial f (\textbf{z}(t),t,\theta)}{\partial \textbf{z}}.
 \end{equation}
 Finally, we compute the gradients with respect to the parameters $\theta$ evaluating a third integral:
 \begin{equation}
     \frac{d\mathcal{F}}{d\theta}=
     \int_{t_1}^{t_0} \textbf{a}(t)^{ \mathsf{T}}
     \frac{\partial f (\textbf{z}(t),t,\theta)}{\partial \theta } dt.
 \end{equation}
All the integrals are evaluated in a single call to the ODE solver, and the Jacobian is computed by automatic differentiation.

The choice to extend the FlowPM code with the ODE implementation is motivated by the fact that to compute the gradient of the forward model, the original algorithm needs to store all the intermediate steps of the simulations. This induces a memory overhead that scales with the number of time steps in the simulation. In the adjoint ODE approach, this is instead replaced by solving another ODE backward in time when evaluating the gradient. 
We illustrate the potential of differentiating through ODE solvers, highlighting the fact that the simulations and the gradients presented in this paper are computed using one single GPU for $128^3$ particles.

\subsubsection{ Hybrid Physical-Neural ODE}\label{N_body_des}    
PM simulations can be used as a viable alternative to full N-body to model the galaxy statistics and create fast realizations of large-scale structures at lower computational cost. Nevertheless, these kinds of simulations lack resolution on small scales and are not able to resolve structures with scales smaller than the mesh resolution.
To compensate for
the small-scale approximations and recover the missing power, we adopt an Hybrid Physical-Neural (HPN) approach presented in \citet{lanzieri2022hybrid}. 
The correction scheme we implement consists in computing the short-range interaction as an additional force parameterized by a Fourier-space Neural Network.
This residual force is modeled by applying a learned isotropic Fourier filter acting
on the PM-estimated gravitational potential $\phi_{PM}$:
\begin{equation}\label{hybrid_model}
    F_\theta(\mathbf{x}, a) = \frac{3 \Omega_m}{2}  \nabla \left[ \phi_{PM} (\mathbf{x}) \ast \mathcal{F}^{-1}(1 + f_\theta(a,|\mathbf{k}|)) \right],
\end{equation}
where $\mathcal{F}^{-1}$ is the inverse Fourier transform and $f_\theta(a,|\mathbf{k}|)$ is a B-spline function whose coefficients are the output of the Neural Network of parameters $\theta$ trained using the CAMELS simulations \citep{villaescusa2021camels}. In particular,  we use a \textit{single} CAMELS N-body simulation at the fiducial cosmology of $h=0.6711$, $n_s =0.9624$, $M_{\nu} = 0.0$ eV, $w=-1$, $\Omega_k = 0.$, $\Omega_m=0.30$, $\sigma_8= 0.8$.

We adopt a loss function penalizing both the positions
of the particles and the overall matter power spectrum at
different snapshot times s:
\begin{equation}\label{loss2}
    \mathcal{L} =  \sum_{s} || \mathbf{x}^{Camels}_s - \mathbf{x}_s ||_2^2  + \lambda || \frac{P_s(k)}{P_s^{Camels}(k)} -1 ||_2^2 \; ,
\end{equation}
where $\lambda$ is an hyper-parameter balancing the
contributions of the two terms. By comparing the results obtained from different values of $\lambda$ in the fiducial setting and outside the training regime, we find that $\lambda= 0.1$ provided the optimal balance in terms of overall correction and overfit.

In \citet{lanzieri2022hybrid} we tested the robustness of the HPN correction scheme to changes in resolution and cosmological parameters, i.e. we applied the correction parameters found with the setting described above to simulations of larger volume or different $\Omega_m$ and $\sigma_8$. We observed that most of the missing power that characterizes the matter power spectrum in the PM approximation is still recovered when the HPN correction is applied to simulations different from the ones used to train the Neural Network. So, from these tests, we can conclude that the HPN is still robust to differences among simulation settings.

\subsection{Differentiable Lensing Simulations}
To extract the lens planes and construct the lightcone, we export 11 intermediate states from the N-body simulation of a fixed interval of 205 $h^{-1}$ Mpc in a redshift range between $z=0.03-0.91$. 
To recover the redshift range of the lightcone, one unit box is replicated using periodic boundary conditions. First, we generate rotation matrices along the three axes, hence, each snapshot is rotated around each of the three axes, finally, all the particles are randomly shifted along the axes. 
To obtain the final density field, each snapshot is projected in a 2D plane by estimating its density with a  cloud-in-cell (CiC) interpolation scheme \citep{hockney1988computer}. After creating a Cartesian grid of coordinates, each slice is interpolated onto sky coordinates. This procedure differs from the one implemented in the MADLens package \citep{bohm2021madlens}. In MADLens the lightcone is built by translating the redshift of the particles into distances, then the particles are projected onto the convergence map at the proper evolution step corresponding to that distance.

\subsubsection{Implementation of Born lensing}
We generate the convergence map by integrating the lensing density along the unperturbed line of sight, i.e. applying the Born approximation \citep{schneider2006weak}. In particular we discretize the \autoref{born_approx} such that it becomes:
\begin{equation} 
    \kappa_{born}= \frac{3H_0^2 \Omega_m}{2c^2}
    \sum_i 
    \bar{\delta_i} \left ( 1-\frac{\chi_i}{\chi_s}\right)
    \left (  \frac{\chi_i}{a_i}\right)
     \Delta \chi
\end{equation}
where the $i$ index runs over the different lens planes, the $\bar{\delta_i}$ indicates the matter overdensity projected into the lightcone, $\chi_s$ defines the comoving distance of the source and $\Delta \chi$ is the  width of the lens plane.

\subsubsection{Implementation of IA with NLA }
We model the effect of IA on the convergence map level following the model proposed by  \citet{fluri2019cosmological}. This allows us to create pure IA convergence maps to combine with shear convergence maps in order to generate a contaminated signal. 
Following \cite{harnois2021cosmic}, the Fourier transform of the intrinsic ellipticities can be phrased as: 
\begin{align}\label{trans_ell}
    \tilde{ \epsilon}_{1}^{IA}(\textbf{k}_{\bot}) & \propto
    \left ( \frac{k_x^2-k_y^2}{k^2} \right )  \tilde{\delta}_{2D}(\textbf{k}_{\bot})
   \mathcal{G}_{2D}(\sigma_g) \\
       \tilde{ \epsilon}_{1}^{IA}(\textbf{k}_{\bot}) & \propto
    \left ( \frac{k_x k_y}{k^2} \right )  \tilde{\delta}_{2D}(\textbf{k}_{\bot})
   \mathcal{G}_{2D}(\sigma_g), \nonumber
\end{align}
where $\sigma_g$ defines the smoothing scale of a two-dimensional smoothing kernel $\mathcal{G}_{2D}$, the tilde symbols $\sim $ refers to the Fourier transformed quantities and $\textbf{k}_{\bot}$ denotes the two Fourier wave-vector components perpendicular to the line of sight.
Combing \autoref{complex_e} and \autoref{trans_ell} we can calculate the intrinsic alignment as part of the convergence map:
\begin{equation}\label{IA_maps}
    \kappa_{IA_{i}}= 
    - A_{IA}\bar{C}_1\rho_c \Omega_m
    \int_{z_{min}}^{z_{max}} 
    n_{i}(z) \delta_{s_{i}}
    \frac{ dz}{D(z)}
\end{equation}
where the index \textit{i} refers to the \textit{i}-th redshift bins.

\subsubsection{Differentiable Peak Counts}
One of the difficulties in estimating derivatives of traditional peak count statistics is that it relies on building
an histogram of peak intensities, and histograms, due to the discrete nature of the bins, are not differentiable. However, the underlying idea of peak counting is just to build an estimate of the density distribution of number of peaks as a function of their intensity. Histograms are one way to build such an estimate, and have been historically preferred, but for no particular reason. To circumvent the non-differentiability of histograms, we prefer here to estimate this density using an alternative method: Kernel Density Estimation (KDE). As a continuous equivalent to an histogram, KDEs are differentiables and can just as well be used to define the peak counts statistic.  
We define the KDE for the peak counts as:
\begin{equation}
   \text{KDE}= \frac{1}{ b_w \sqrt{2\pi}} \exp{\left ( -\frac{(X-x)^2}{2 b_w^2} \right )}
\end{equation}
where $b_w$ is the smoothing bandwidth parameter, \textit{X} is the number of peaks in a given bin, and \textit{x} is the center of each bin.

This procedure yields a peak count statistic that is smoothly differentiable with respect to the input map and thus can be used for applications such as Fisher forecasting as discussed later in this paper.

\section{Validating simulations for LSST}\label{Validating_simulations_for_LSST_Y}
In this section, we compare the results from our simulation to other works, including the analytic models for the matter power spectrum \textit{Halofit} \citep{smith2003stable,takahashi2012revising} and the cosmological N-body simulations, $\kappa$TNG \citep{osato2021kappatng}.

The $\kappa$TNG simulations is a suite of publicly available weak lensing mock maps based on the cosmological hydrodynamical simulations IllustrisTNG, generated with the moving mesh code AREPO \citep{2010MNRAS.401..791S}.
In particular, we use the $\kappa$TNG-Dark suit of maps based on the corresponding dark matter-only TNG simulations. 
The simulations have a side length of the box equal to 205 Mpc/h and $2500^3$ CDM particles.
To model the propagation of light rays and simulate the weak lensing maps, a multiple-lens plane approximation is employed. 
The simulation configuration consists of a map size of $5\times5$ deg$^2$, $1024\times 1024$ pixels and a resolution of 0.29 arcmin/pixel. 
For a complete description of the implementation see \cite{osato2021kappatng}.

To produce our simulations, we follow the evolution of $128^3$ dark matter particles in a periodic box of comoving volume equal to $205^3$ ($h^{-1}$ Mpc)$^3$, with initial conditions generated at z=6 using the linear matter power spectrum as implemented by \cite{eisenstein1998baryonic}. In particular, we implement the  Eisenstein-Hu transfer function in the Tensorflow framework, in order to compute its gradients automatically.

We assume the following cosmological parameters: $h=0.6774$, $n_s =0.9667$, $M_{\nu} = 0.0$ eV, $w=-1$, $\Omega_k = 0.$, $\Omega_m=0.3075$, $\sigma_8= 0.8159$, such that they match the results of Planck 2015 \citep{ade2016planck}.
We reproduce the same configuration of $\kappa$TNG, i.e. each map is on a regular grid of $1024^2$ pixels and $5\times 5$ deg$^2$.

The actual choice of bins to include in the forecasting is made following the DESC data requirement for the angular power spectra \citep{mandelbaum2018lsst}, i.e. adopting $\ell_{max,shear} = 3000$ and  $\ell_{min,shear} = 300$.

\subsection{HPN validation }
\begin{figure}
    \centering
    \includegraphics[width=\columnwidth]{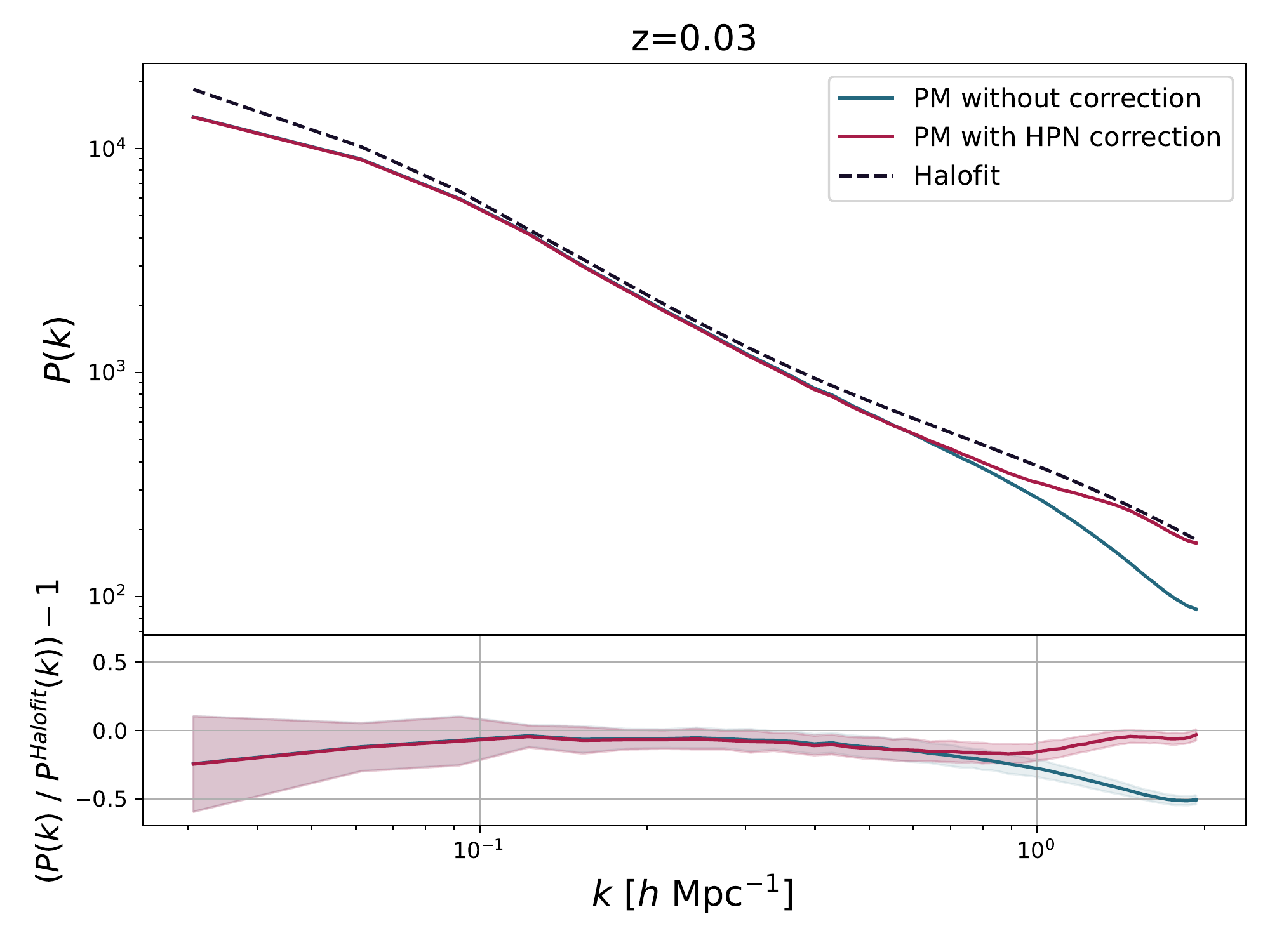}
    \includegraphics[width=\columnwidth]{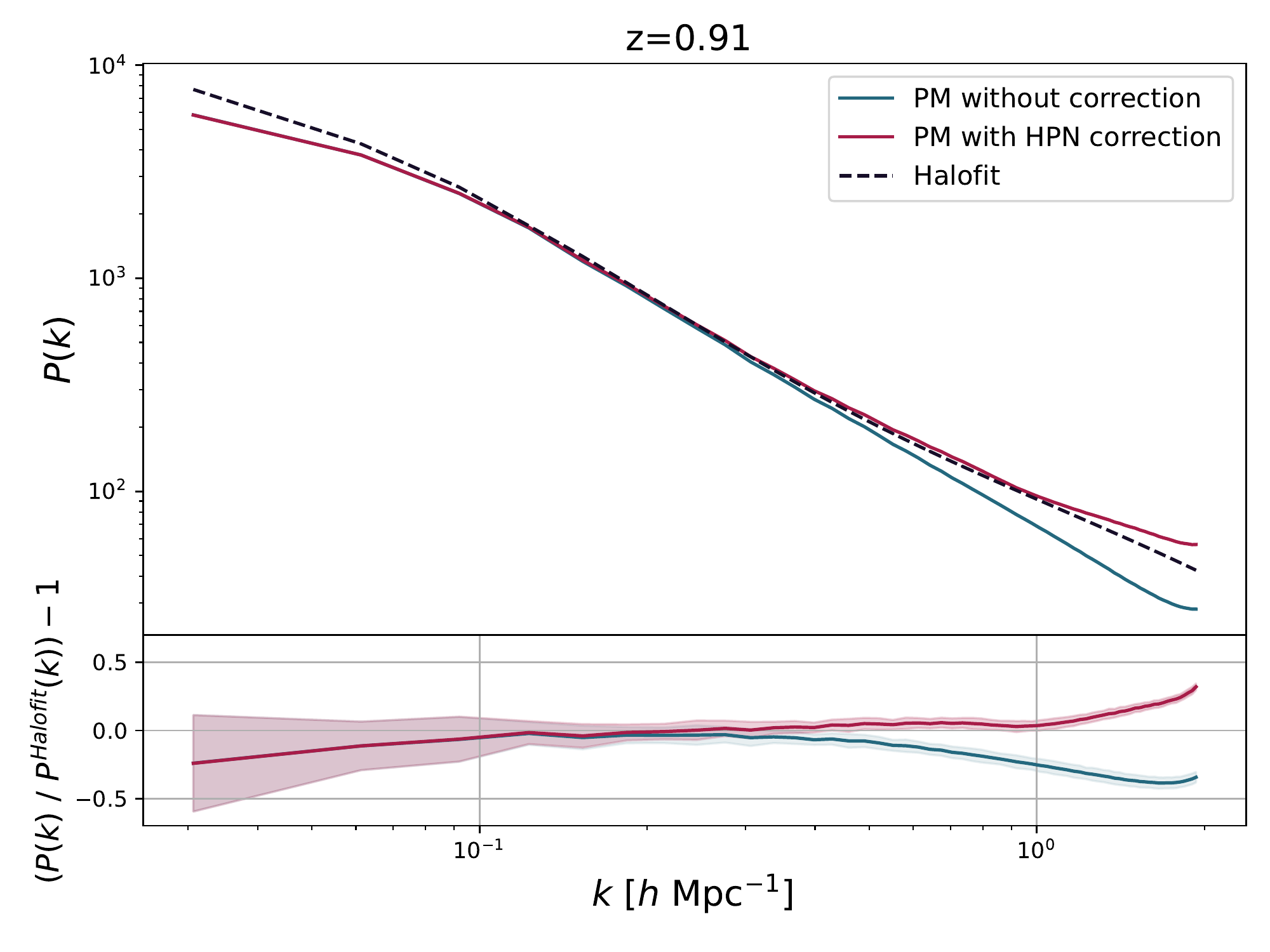}
    \caption{
     Matter power spectrum and fractional matter power spectrum of PM simulations before and after using the Hybrid Physical-Neural (HPN) correction model and the theoretical halofit model for redshift z=0.03 (upper panel) and redshift z=0.91 (lower panel). The power
spectra and ratios are means over 100 independent map realisations. The shaded regions represent the standard deviation from 100 independent DLL realisations.
    }
    \label{fig:pkhalofit_comp}
\end{figure}

To compensate for the small-scale approximations resulting from PM, we applied the HPN approach presented in \autoref{N_body_des}. 
We show on \autoref{fig:pkhalofit_comp} the power spectrum  and the fractional power spectrum of PM simulations before and after the HPN correction compared to analytic Halofit predictions \citep{smith2003stable, takahashi2012revising} for redshift $z=0.03$ and $z=0.91$.
 We observe a bias between our measured power spectrum and the theoretical prediction at low $k$. This reduced power is explained by the small box size of our simulation and the associated reduced number of large-scale modes.  At redshift $z=0.91$ most of the missing
power is recovered by the HPN correction up to $k\sim 1$, after which the method overemphasizes the small-scale power. In this article however, we can assume that this effect does not impact the results of the cosmological parameters forecast, since it concerns scales that are beyond the range of frequencies that are taken into account for the analysis.

At redshift $z=0.03$, the correction model does not improve significantly the results.

\begin{figure*}
    \centering
    \includegraphics[width=\textwidth]{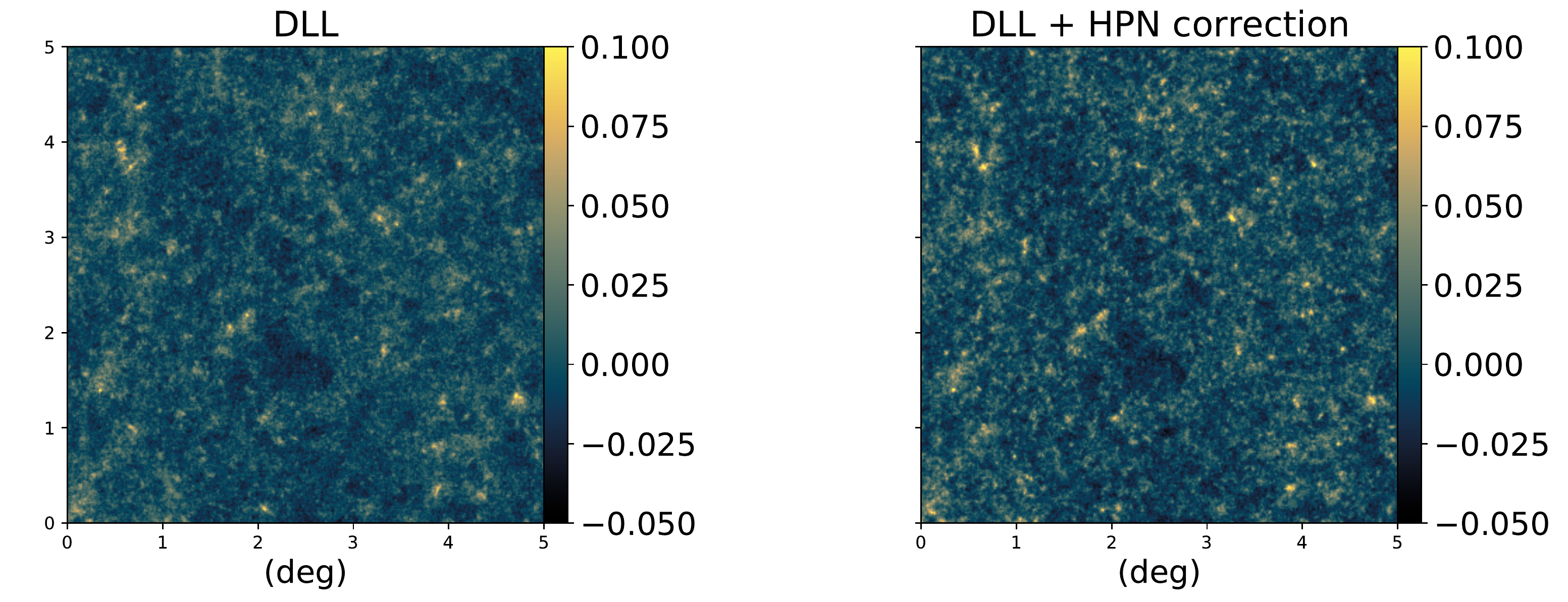}
    \caption{ \textbf{Left panel}: Convergence map at source redshift $z= 0.91$ from DLL, PM only.
    \textbf{Right panel}: Same convergence map when the HPN correction is applied. 
    }
    \label{fig:comp_kmap}
\end{figure*}

In \autoref{fig:comp_kmap} we show an example of our convergence map at $ z= 0.91$, from pure PM simulation (first panel) and the HPN corrected simulation (second panel). The HPN model sharpens structures in the lensing field without introducing any artifacts.

\begin{figure}
    \centering
    \includegraphics[width=0.95\columnwidth]{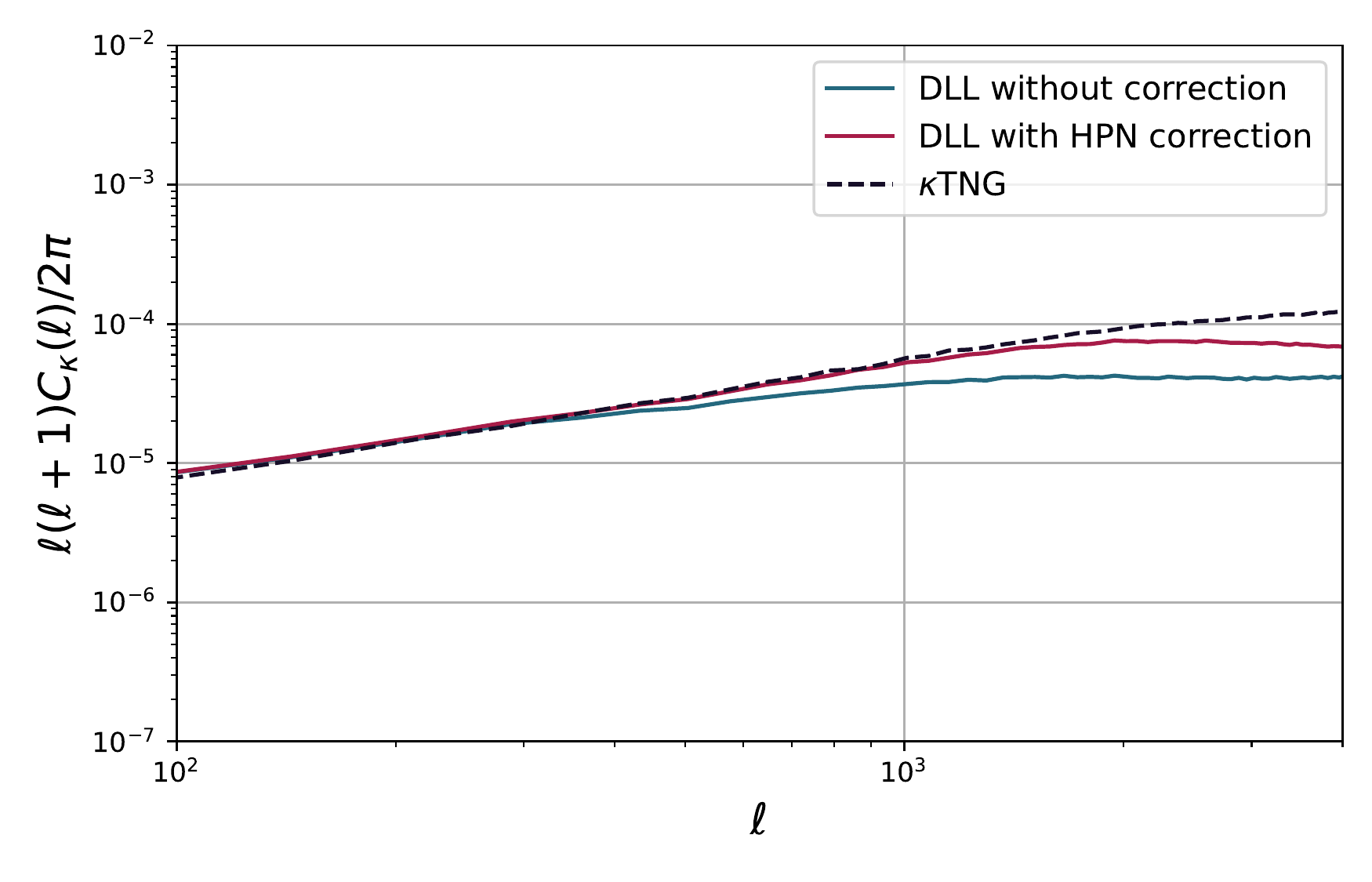}
        \includegraphics[width=\columnwidth]{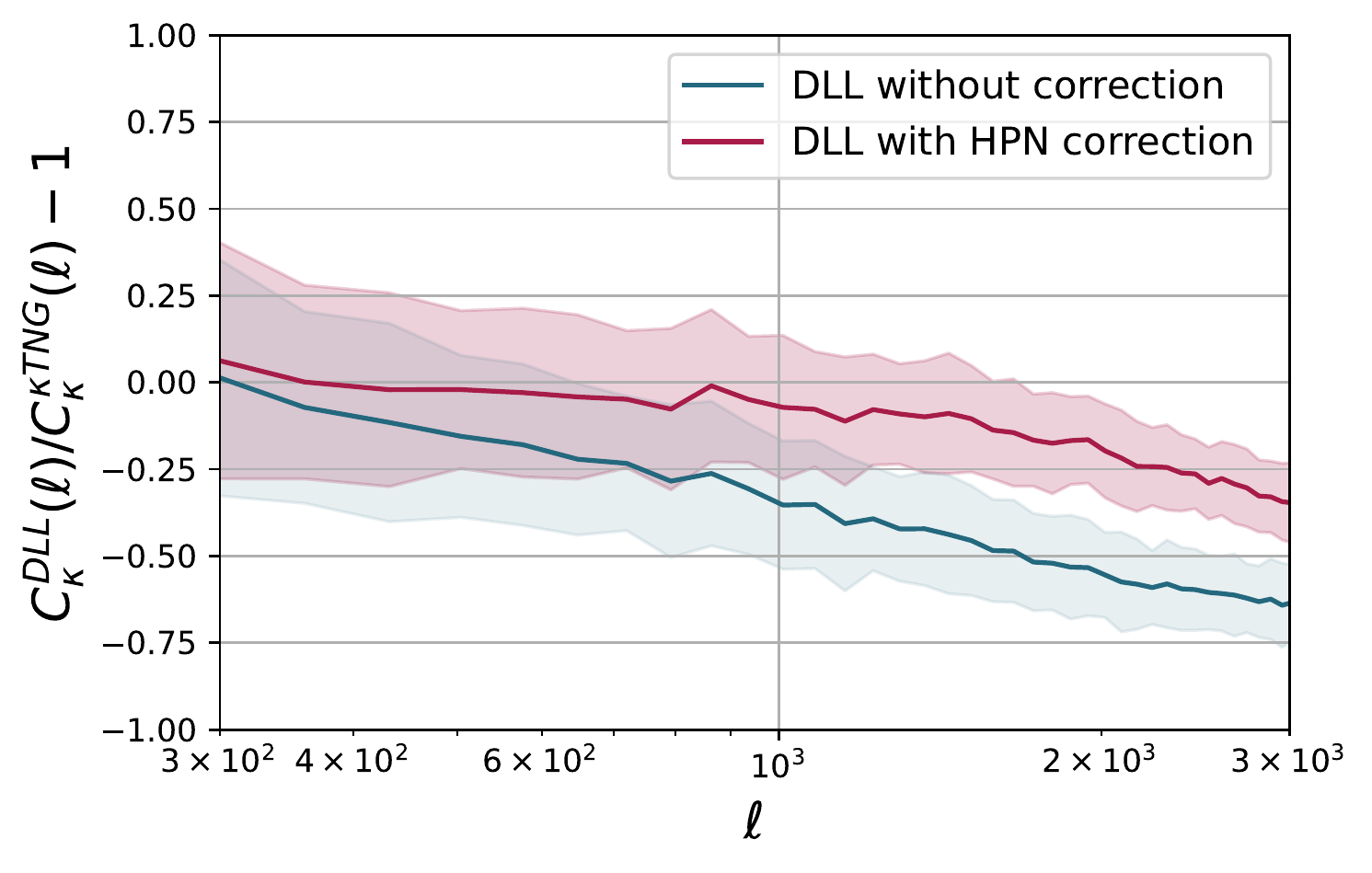}
    \caption{
    \textbf{Upper panel}: Angular power spectra of PM simulations before and after using the Hybrid Physical-Neural (HPN) correction model compared to the $\kappa$TNG prediction. \textbf{Lower panel}: fractional angular power spectrum of PM simulations before and after using the Hybrid Physical-Neural (HPN) correction model and the $\kappa$TNG prediction. The power spectra and ratios are means over 100 independent map realisations and the shaded regions represent the standard deviation from 100 DLL realisations. The spectra are computed for the source redshift $z_s=0.91$. 
}
    \label{fig:ps_comp}
\end{figure}

In the upper panel of \autoref{fig:ps_comp}, we present the angular power spectrum computed from our Differentiable Lensing Lightcone (DLL hereafter) complemented by the HPN scheme and a conventional DLL simulation with the same resolution. Both the outputs are compared to the $\kappa$TNG prediction. In the lower panel  of \autoref{fig:ps_comp} the fractional differences between the convergence power spectra from the two maps and the $\kappa$TNG are shown.  Both the power spectra and ratios are averaged over $N = 100$ realisations. 
We can see that the HPN  model reduces the relative deviations of the angular power spectra to within 30\%. 
We also observe a perfect match at large scales, since the $\kappa$TNG and the DLL simulations have the same box size of 205 Mpc/h$^3$.

\subsection{IA validation}
\begin{figure}
    \centering
    \includegraphics[width=0.95\columnwidth]{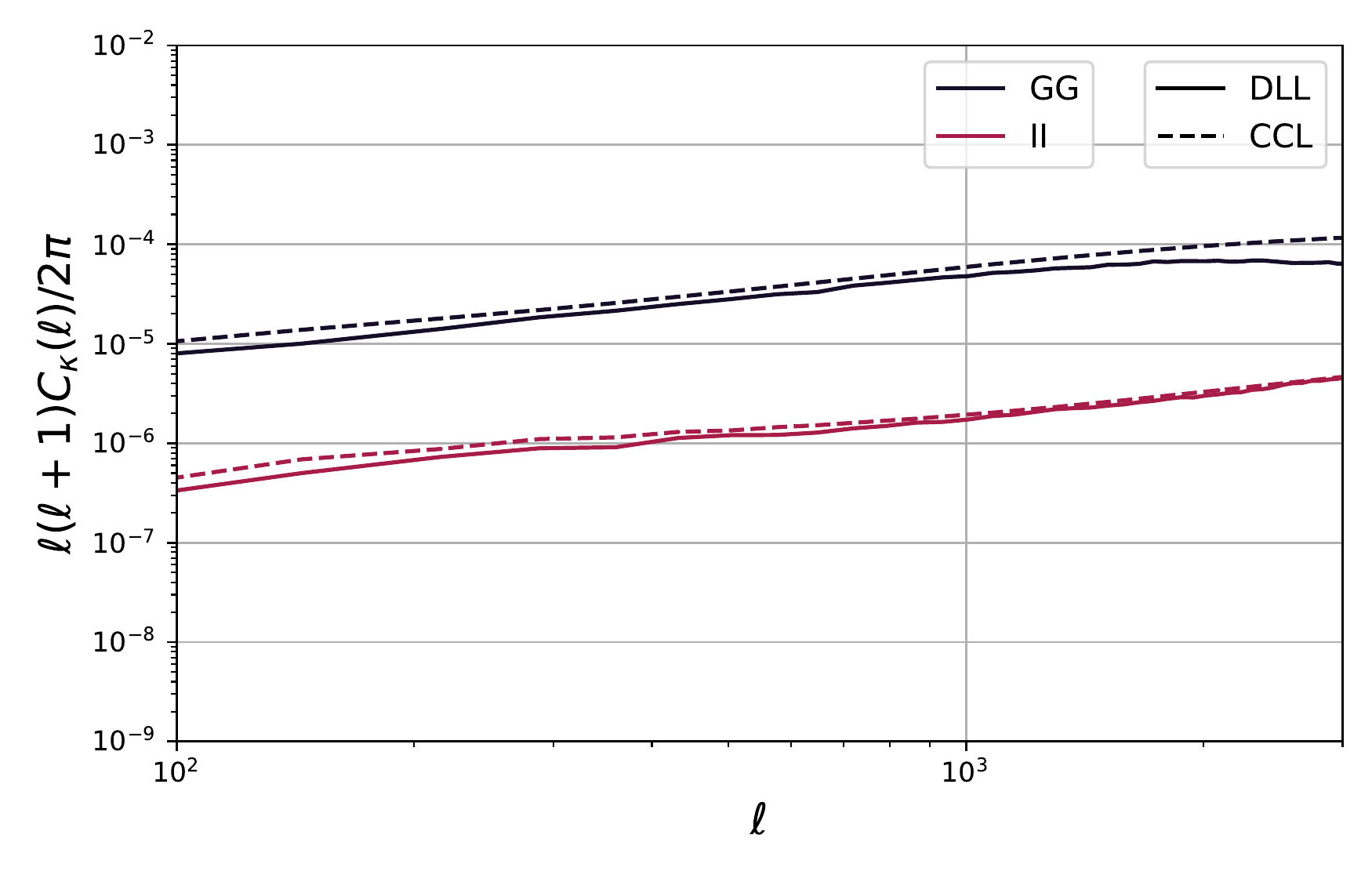}
    \includegraphics[width=\columnwidth]{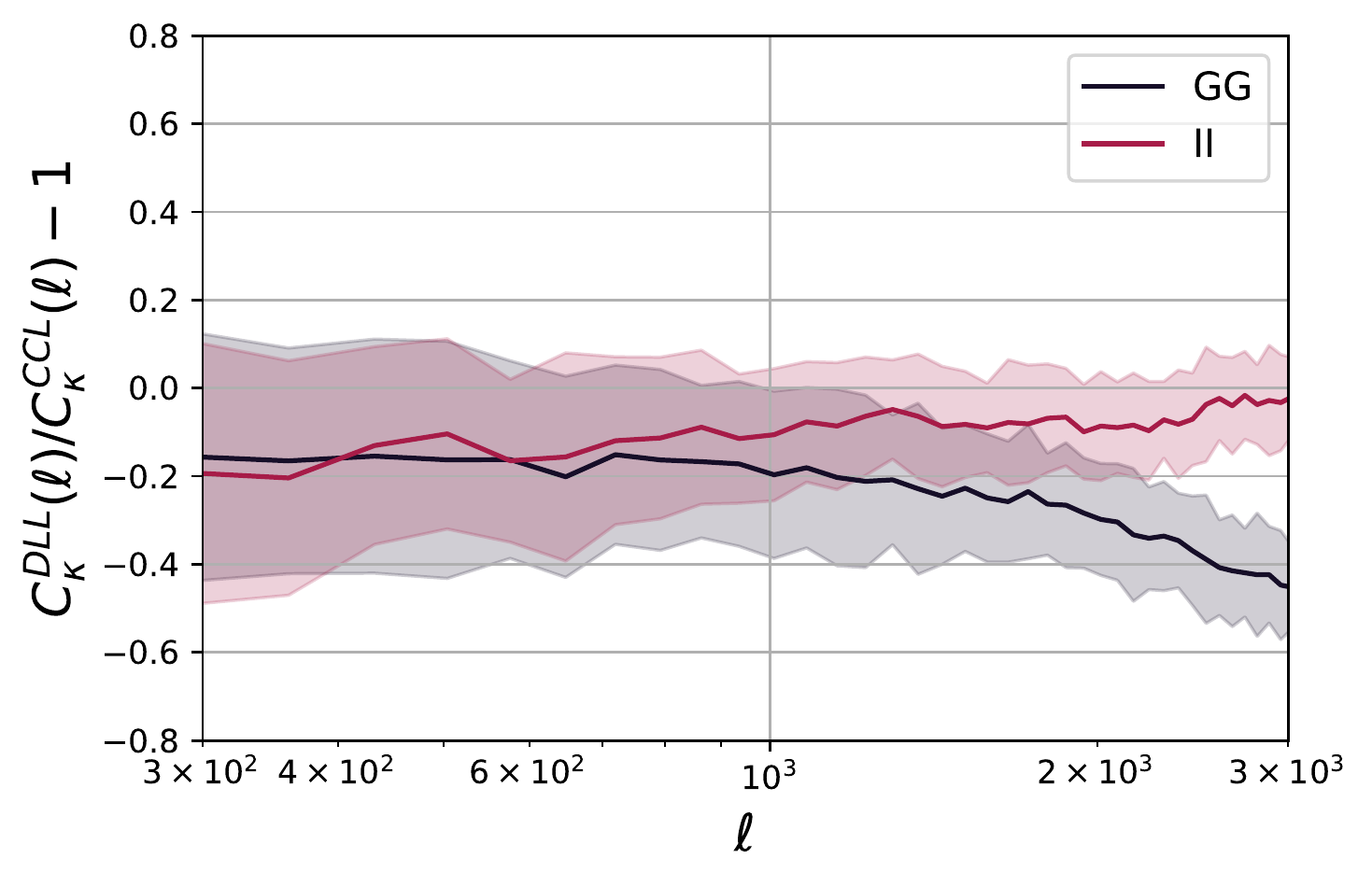}
    \caption{ \textbf{Upper panel}: The $C_{\ell}^{II}$ and $C_{\ell}^{GG}$ contributions from theoretical predictions (dashed line) and DLL simulations. \\
     \textbf{Lower panel}: The fractional difference between the theoretical and simulated $C_{\ell}^{II}$ and $C_{\ell}^{GG}$ contributes.
     We can see that we measure a reduced power spectrum at low $\ell$ compared to the theoretical predictions. This can be explained by the small volume of our simulation and the related low number of large-scale modes.
     The power spectra and ratios are means over 100 independent map realisations and the shaded regions represent the standard deviation from 100 realisations.}
    \label{fig:ia_validation}
\end{figure}

In the upper panel of  \autoref{fig:ia_validation}, we present the $C_{\ell}^{II}$ and $C_{\ell}^{GG}$ contributes from our DLL simulations compared to theoretical Halofit predictions \citep{smith2003stable}. In the lower panel of the same figure, we show the fractional differences between the mentioned contributions.
To validate the IA infusion, only for this experiment, we run simulations keeping the term $A_{IA}=1$. 
As we can see, the fractional difference for the $C_{\ell}^{II}$ term features uncertainty consistent with $C_{\ell}^{GG}$ term, validating our infusion process. 
The signal is computed for the source redshift $z_s=0.91$ and is averaged over 100 realizations.
 The theoretical predictions are computed using the public \href{https://ccl.readthedocs.io/en/latest/}{Core Cosmological Library} (CCL, \citet{chisari2019core}). 

\subsection{Lensing $C_\ell$}
\begin{figure}
    \centering
    \includegraphics[width=0.95\columnwidth]{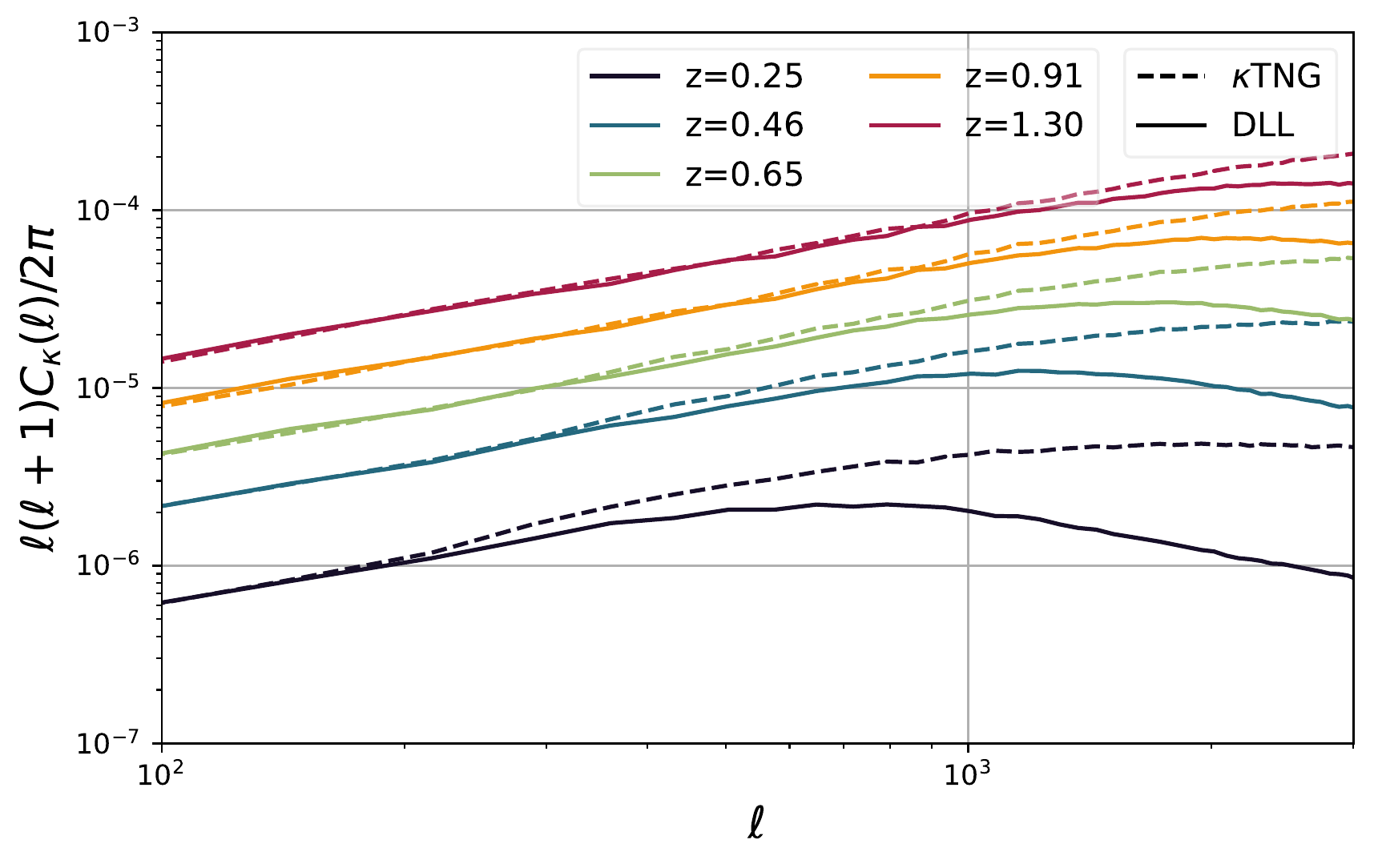}
    \includegraphics[width=\columnwidth]{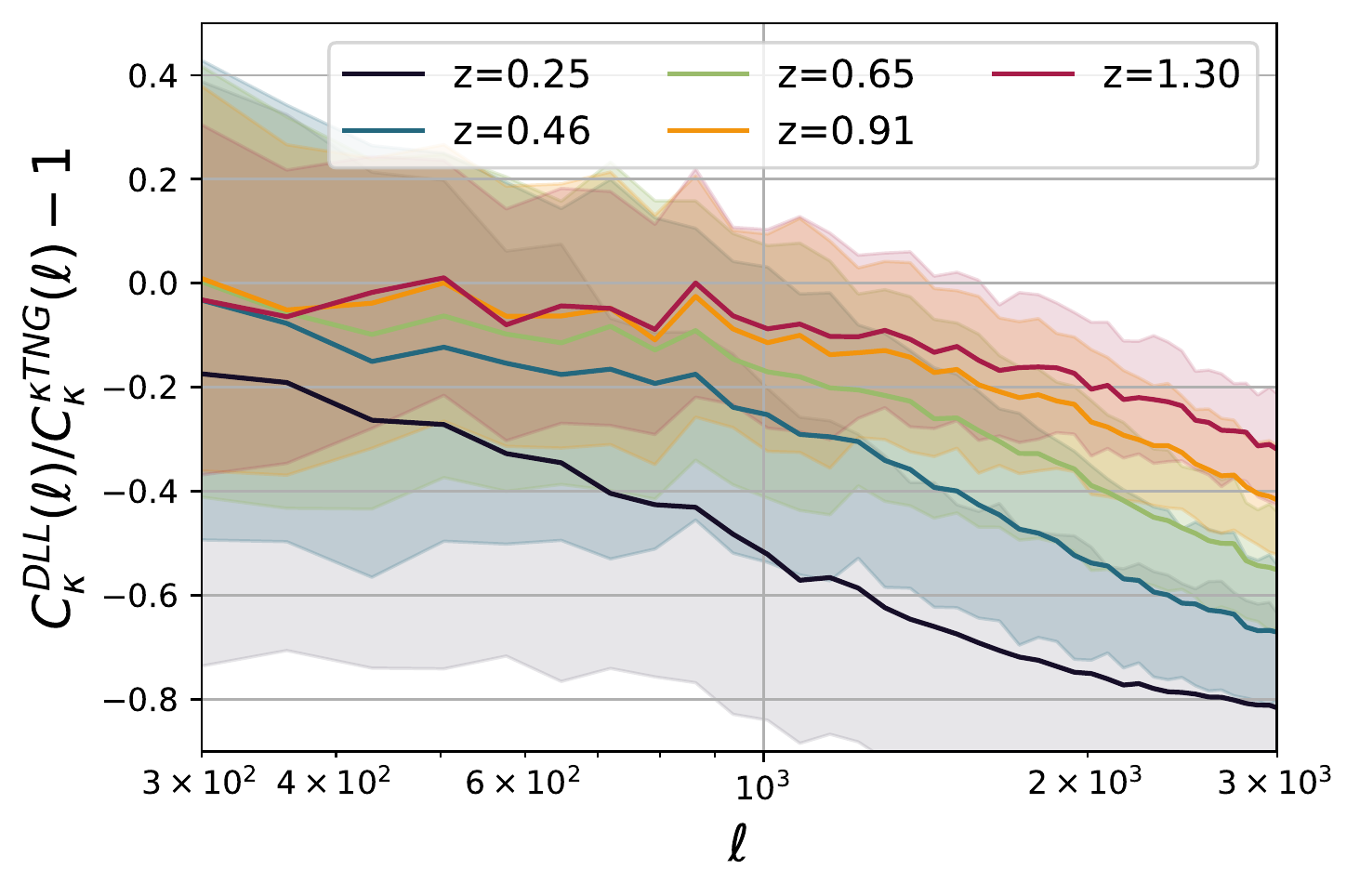}
    \caption{
    \textbf{Upper panel}: Angular power spectra for 5 source redshift from
our DLL maps compared to the $\kappa$TNG predictions.
 \textbf{Lower panel:} Fractional angular power spectra of DLL simulations and $\kappa$TNG simulations for different source redshift.
  The power spectra mean over 100 independent map realisations and the shaded regions represent the standard deviation from 100 independent DLL realisations.}
     \label{fig:clsktng_comp}
\end{figure}

To quantify the accuracy of the simulations we aim to reproduce the summary statistics from the Dark Matter Only $\kappa$TNG simulations. 
We compare the results from the angular power spectrum for different source redshift, just investigating how well we can recover the power spectrum for a given source plane. The results of the angular power spectrum from the sources redshift z=[0.25,0.46,0.65,0.91,1.30] are shown in the upper panel of \autoref{fig:clsktng_comp}, as well as the fractional differences between the $\kappa$TNG and DLL maps in the lower panel.
We observe that the differences for $z_s=0.91$ and  $z_s=1.30$ curves are within $10\%$ of accuracy for scales larger than $\ell=1000$, within $25\%$  for scales $1000<\ell<2000$ and  within $25\%$ and $45\%$ for scales $2000<\ell<3000$. For lower source redshifts the deficit of power in our simulations becomes worse. This can be explained considering that a given value of $\ell$ at lower redshift corresponds to smaller scales, some of those below the resolution of our simulations. We conclude that, if for z=0.91 and z=1.30 we have a general agreement with $\kappa$TNG, with this specific setting of the model, we can not model correctly cases with sources redshift lower than z=0.91.

 We want to highlight that the results shown are produced keeping the resolution of the simulations extremely low, and we do not aspect to get the same precision as $\kappa$TNG. The purpose of these tests, and the overall goal of the paper, is to present a proof of concept of the DLL package and its potential.  In practice, we will not work at this resolution. 

Nevertheless, note that the simulations presented here already achieve a similar resolution of the MassiveNus simulations \citep{liu2018massivenus}, despite being generated using one single GPU.

\subsection{Peak counts}\label{peak_sec}
We compute the starlet peak counts as wavelet coefficients with values higher than their eight neighbors. 
We define $J_{max}=7$ in \autoref{wav_des}, this starlet filter applied to our map with a pixel size of 0.29 arcmin, corresponds to a decomposition in 7 maps of resolution [0.59, 1.17, 2.34, 4.68, 9.33, 18.79, 37.38] arcmin and a coarse map. To satisfy the survey requirement and keep the analysis centred in the range $\ell=[300, 3000]$, we consider only the scales corresponding to [9.33, 18.79, 37.38] arcmin. The peaks are counted for 8 linearly spaced bins within the range $(\kappa*\mathcal{W})=[-0.1, 1.]$. 

\begin{figure}\label{peaks_z}
    \centering
    \includegraphics[width=\columnwidth]{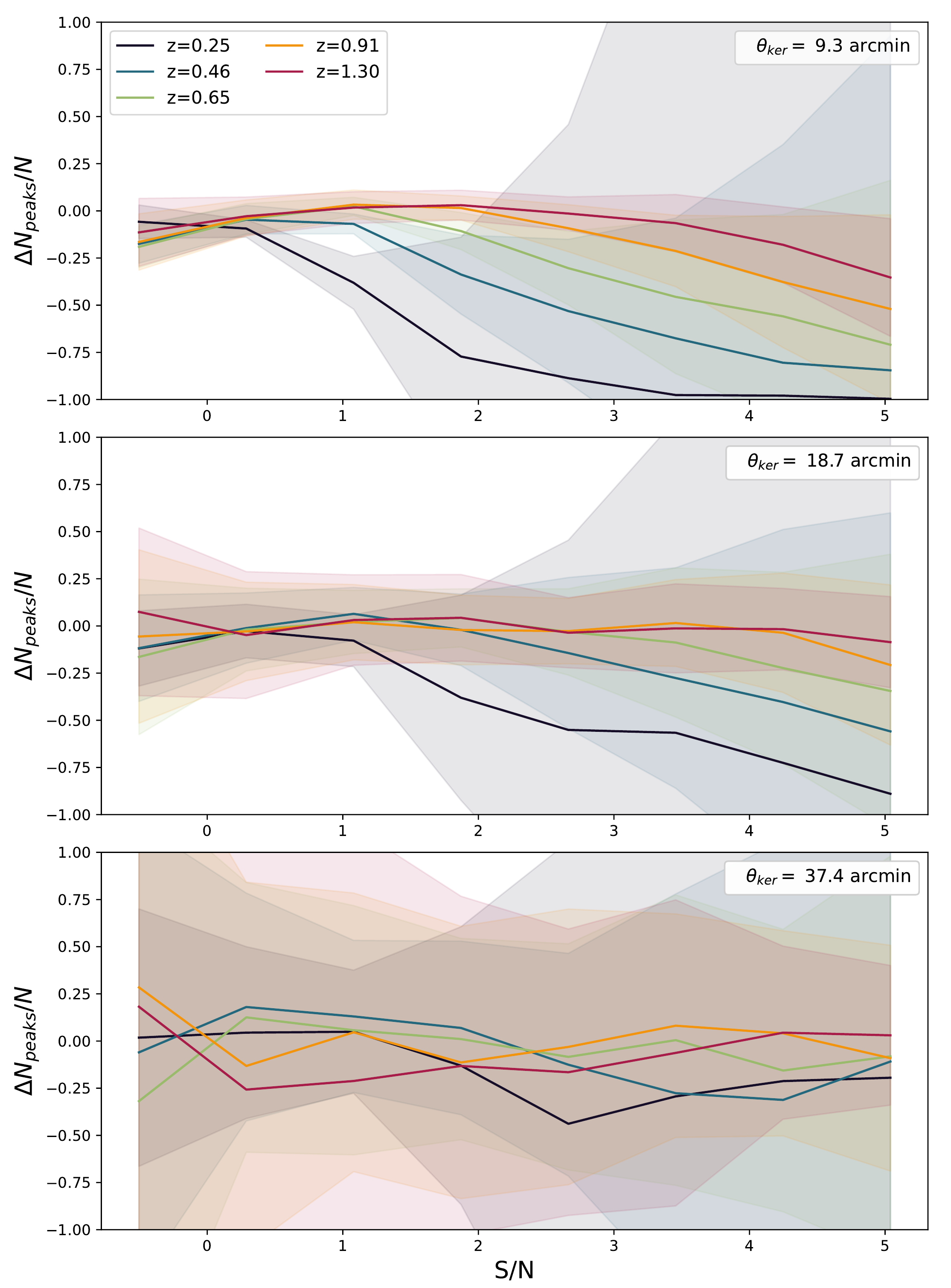}
    \caption{Fractional number of peaks of DLL simulations and $\kappa$TNG simulations for different sources redshift. The peak counts distributions are shown for each starlet scales resolutions used: 9.34 (upper panel), 18.17 (center panel), 37.38 arcmins (lower panel).
  The results mean over 100 independent map realisations and the shaded regions represent the standard deviation from 100 independent DLL realisations.}
\end{figure}

As for the power spectrum, we compare the peak counts statistic from our map to the one from the $\kappa$TNG for different redshift bins. We present the results in \autoref{peaks_z}. These results are shown in terms of S/N, where the signal to noise is defined as the ratio between the amplitude of wavelet coefficients over the noise expected for our survey choice.
At wavelet scale $\theta=9.33$ arcmin the differences for the $z_s=0.91$ curves are within the $20\%$ up to $S/N=3$, for $S/N>3$ the accuracy is between the $20\%$ and the $50\%$. At larger scale, $\theta=18.79$ arcmin the accuracy is within the $20\%$. Finally, at $\theta=37.38$ arcmin the accuracy is within the $15\%$, except $S/N<1$ where the accuracy decreases up to $28\%$.
The results slightly improve for z=1.30, showing an accuracy within the $35\%$ for scale $\theta=9.33$ arcmin,  within the $10\%$ for $\theta=18.79$ arcmin and $25\%$ for $\theta=37.38$ arcmin.
 As for the power spectrum case, we observe higher discrepancies at lower redshift, hence we can conclude that, with the current setting of our simulation, we can not model correctly such redshift.

\section{Application: Fisher forecast
}\label{Application_Fisher_forecast_for_HOS}
As an example of application of differentiable simulations, we aim to investigate the degeneracy between the cosmological parameters in high dimensional space and when systematics, such as the intrinsic alignment, are included in the analysis. Thanks to automatic differentiation, 
taking the derivative through the simulation with respect to the initial cosmological and nuisance parameters is now possible, thus allowing, among other things, for Fisher forecasts.
In this section, we briefly introduce the Fisher forecast formalism. We also describe in detail the specific choices for the analysis we use throughout.

\subsection{Forecast formalism}
Fisher forecast is a widely used tool in cosmology for different purposes, e.g. investigate the impact of systematic sources or forecast the expected constraining power of the analysis \citep{tegmark1997karhunen}. It can be thought as a tool to forecast error from a given experimental setup and quantify how much information we can extract from it. 
The Fisher matrix is defined as the expectation value of the Hessian matrix of the negative log-likelihood $\mathcal{L}(C(\ell);\theta)$:
\begin{equation}
    F_{\alpha \beta}=\left \langle 
    \frac{\partial^2\mathcal{L}}
    {\partial \theta_{\alpha}\partial \theta_{\beta}}
    \right \rangle \;,
\end{equation}
where we indicate with $\theta$ the cosmological parameters or any systematics included in the simulation.
If we assume a Gaussian likelihood and a Covariance matrix $C_{ij}$ independent from $\theta$, we can calculate the Fisher matrix as
\begin{equation}\label{Fisher_matrix}
      F_{\alpha \beta}= \sum_{ij}\frac{\partial \mu}{\partial \theta_{\alpha}} C_{ij}^{-1}
      \frac{\partial \mu}{\partial \theta_{\beta}}
\end{equation}
were we indicate as $ \frac{\partial \mu}{\partial \theta_{\alpha}} $ the derivatives of the summary statistics w.r.t the cosmological or nuisance parameters evaluated at the fiducial values. So, under the assumption of Gaussian likelihood, the Fisher information matrix provides a lower bound on the expected errors on cosmological parameters. 

\subsection{Analysis choices} \label{Analysis_choices}

\begin{figure}\label{window}
    \centering
    \includegraphics[width=0.95\columnwidth]{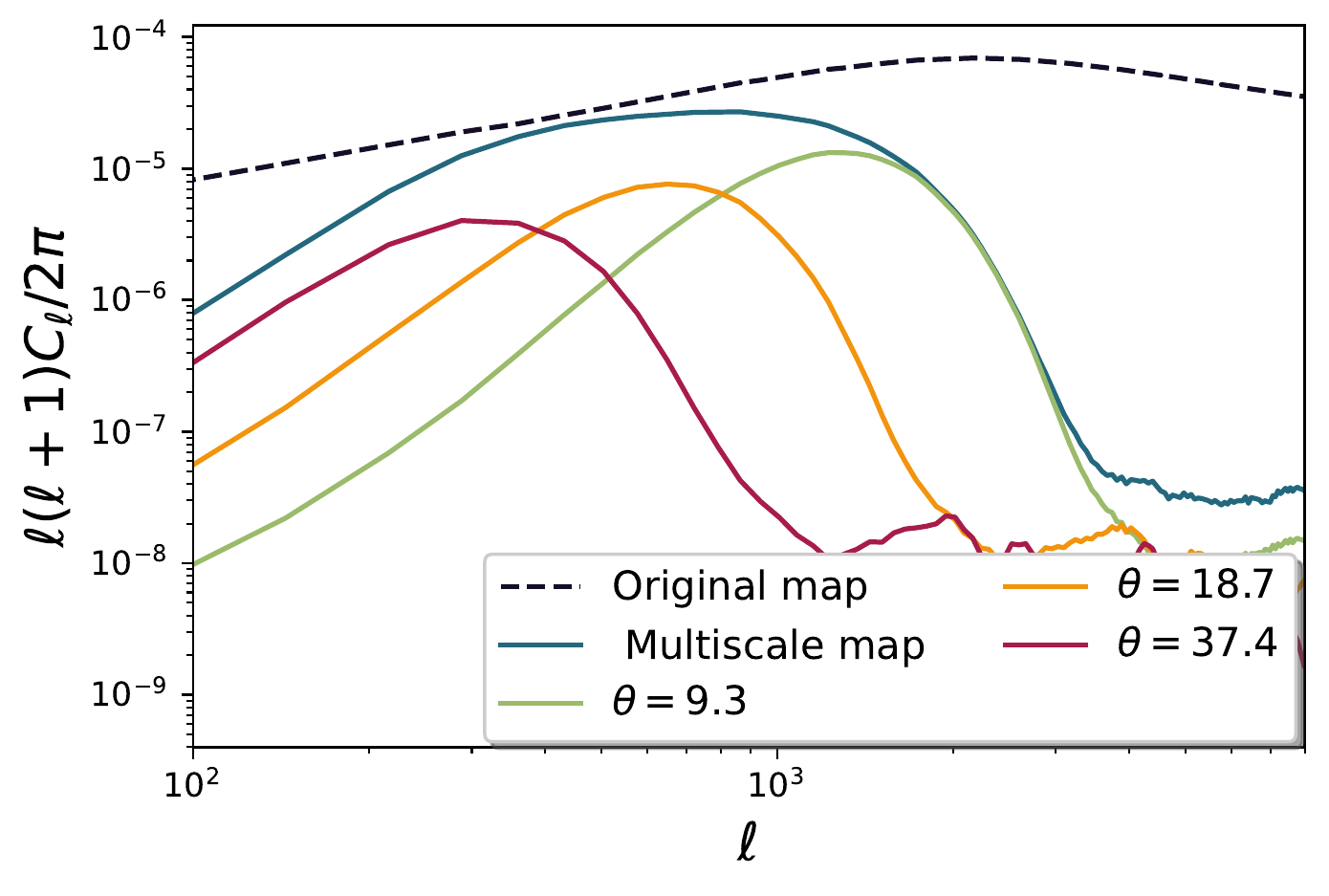}
    \caption{Example of the filtered $C_{\ell}$ used for the analysis. The colored lines show the $C_{\ell}$ computed on maps with a different resolution of the starlet decomposition. Specifically: the blue line (multiscale map) corresponds to $C_{\ell}$ computed on the summed image, the black dashed line (Original map) corresponds to the standard $C_{\ell}$ computed from a non-filtered map. }
     \label{fig:window}
\end{figure}

To perform our study we use a single source redshift at z=0.91. Specifically, we generate 5000 independent map realisations to which we add Gaussian noise with mean zero and variance:
 \begin{equation}
     \sigma^2_n= \frac{\sigma_e^2}{A_{pix}n_{gal}},
 \end{equation}
where we set the shape noise $\sigma_e = 0.26$, the pixel area $A_{pix}=0.29$ arcmin$^2$ and the galaxy number density $n_{gal}=20$ arcmin$^{-2}$.  
We assume a parameter-independent covariance matrix computed as:
\begin{equation}\label{covariance_matrix}
    C_{ij}=\sum_{r=1}^N
    \frac{(x_i^r-\mu_i)(x_j^r-\mu_j)}{N-1}
\end{equation}
where $N$ is the number of independent realizations, $x_i^r$ is the value of the summary statistics in the $i^{th}$ bin for a given realization $r$, and $\mu$ is the mean of the summary statistics over all the realization in a given bin. 
In addition, we adopt the estimator introduced by \cite{hartlap2007your} to take into account the loss of information caused by the finite numbers of bins and realizations, i.e. we compute the inverse of the covariance matrix as :
\begin{equation}
    C^{-1}=\frac{N-n_{bins}-2}{N-1}C_{*}^{-1},
\end{equation}
where $C_{*}$ is the covariance matrix defined in \autoref{covariance_matrix}. 
As mentioned before, we want to focus on a fair comparison between the power spectrum and the peak counts method. To be sure we are considering the same scales for both statistics, we apply a wavelet pass-band filter to the maps to isolate particular scales before measuring the power spectrum.  We use the same scales used for the Peak counts, i.e. we decompose the noisy convergence map in seven images, we sum back only the three maps corresponding to [9.33, 18.79, 37.38] arcmin and compute the angular $C_{\ell}$ on the resulting image.
An example of the $C_{\ell}$ computed for each individual starlet scale image and for the summed image is depicted in \autoref{window}.

For each map, we compute the angular power spectrum and the peak counts by using our own differentiable code implemented in the TensorFlow framework. \footnote{
Code publicly available at: \\ \href{https://github.com/LSSTDESC/DifferentiableHOS/tree/main/DifferentiableHOS/statistics}{https://github.com/LSSTDESC/DifferentiableHOS/statistics}  }

The derivatives with respect to all parameters are evaluated at the fiducial cosmology as the mean of 1500 and 2600 independent measurements for the peak counts and the $C_{\ell}$ respectively. Indeed, while the peak counts reach the convergence with $\sim$ 1500 simulations, the $C_{\ell}$ proves to be more sensitive to noise and thus, requires more realizations to convergence.
In \autoref{fig:Fisher_stab} in appendix B we test the stability of the Fisher contours by changing the number of simulated maps used to compute the jacobian.

The priors used in the forecast process, are listed in \autoref{tab:prior} following \citet{zhang2022transitioning}.
To take into account the partial coverage of the sky, we scale the Fisher matrix by the ratio $f_{\text{map}}/f_{\text{survey}}$, where $f_{\text{map}}$ is the angular extend of our $\kappa$map $f_{\text{map}}=25$ deg$^2$ and $f_{\text{survey}}$ corresponds to the size of the convergence maps for Stage IV-like survey $f_{\text{survey}}=15000$ deg$^2$.

\subsection{Results} 
We now compare the relative constraining power of the two statistics described in \autoref{stat} using the Fisher matrix formalism. 
As mentioned before, our interest is to investigate the sensitivity of the two weak-lensing statistics when systematic, such as the intrinsic alignment, and more cosmological parameters are included in the forecast. 
The results presented here are obtained from one single source redshift at z=0.91, assuming the fiducial cosmology and survey requirement presented in \autoref{Validating_simulations_for_LSST_Y} and \autoref{Analysis_choices}. The fiducial and priors ranges of the parameters are listed in \autoref{tab:prior}.

\begin{figure*}
    \centering
    \includegraphics[width=\textwidth]{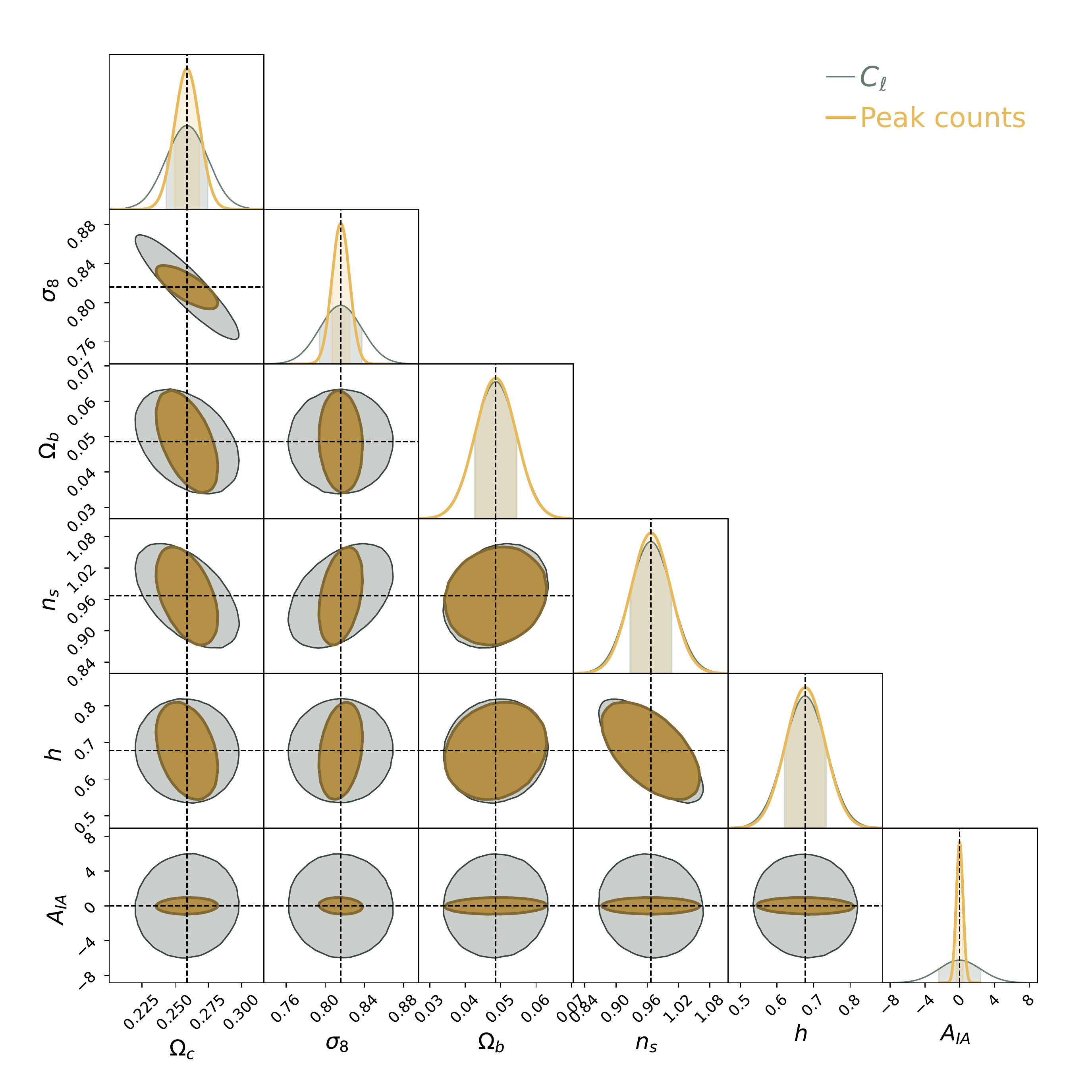}
    \caption{$2\sigma$ contours derived for one single source redshift at z=0.91 and the survey setup presented in \autoref{Validating_simulations_for_LSST_Y}. The constraints are obtained by applying the starlet Peak counts (yellow contours) computed on noisy maps filtered using a starlet kernel of [$9.33, 18.70, 37.38$] arcmin together and the wavelet pass-band filter for the $C_{\ell}$ statistics (grey contours) as described in \autoref{Validating_simulations_for_LSST_Y}.  
The dashed black lines are located at the fiducial parameter values.}
     \label{fig:Fisher_result}
\end{figure*}

\autoref{fig:Fisher_result} shows the $2\sigma$ contours on the full $\Lambda$CDM parameter space and intrinsic alignment term considered for the two analyses. 
The contours obtained by the angular $C_{\ell}$ analysis are plotted in grey, the ones for the peak counts in yellow. 
We find that in constraining $\Omega_c$, $\sigma_8$ and $A_{IA}$ peak counts outperform the power spectrum, while $h$, $n_s$ and $\Omega_b$ parameters, within the limit of our setting, are not constrained by either and are prior dominated.

This is an interesting result, confirming the higher constraining power of weak-lensing peak counts as found in \cite{ajani2020constraining}, especially considering that the two studies differ in multiple aspects. 
The most important difference between these two analyses is the parameters they include. Whereas \cite{ajani2020constraining} derive constraints on the sum of neutrino masses $M_{\nu}$, the total matter density $\Omega_m$, and the primordial power spectrum normalization $A_s$, we include the five cosmological parameters of the $\Lambda$CDM model and intrinsic alignment amplitude. 
The constraining power of the peak count statistic keeps being higher even in high dimensional cosmological parameter space and when the intrinsic alignment is included. 

The chosen angular scales differ as well. \cite{ajani2020constraining} consider angular scales in the range $l = [300,5000]$, while we focus, for both multiscale peak counts and $C_{\ell}$, on scale approximately corresponding to the range $l = [300,3000]$. Despite we are neglecting scales $\ell>3000$, containing a larger amounts of
non-Gaussian information, we find that for mildly non-linear scales we are considering, the peak counts statistic still constrains the cosmological parameters the most.

\begin{table}
	\centering
	\caption{ Prior and fiducial values used for the forecasting.}
	\begin{tabular}{lcc} 
		\hline \hline
		Parameter  & Prior & Fiducial value \\
		$\Omega_c$ & $\mathcal{N}$ (0.2589,0.2) & 0.2589 \\
		$\Omega_b$ & $\mathcal{N}$ (0.0486,0.006) & 0.0486 \\
		$\sigma_8$ & $\mathcal{N}$ (0.8159,0.14) & 0.8159 \\
		h & $\mathcal{N}$ (0.6774,0.063) & 0.6774\\
		$n_s$ & $\mathcal{N}$ (0.9667,0.08) & 0.9667 \\
		$A_{IA}$ &  $\mathcal{N}$ (0,3) &  0.0 \\
		\hline
	\end{tabular}
	\label{tab:prior}
\end{table}
Finally, we find that the contours on the galaxy intrinsic alignment are significantly better constrained by the peak counts.

\section{Discussion}\label{Discussion}
In this section, we discuss the limitations of the methodology and results obtained in this paper highlighting in particular strategies for future extensions and  applications.

In this work, we only used a single source plane in our Fisher forecast analysis, which does not allow us to evaluate the full impact that IA would have in a tomographic analysis. In particular, we do not have a contribution from the GI term. Many studies have demonstrated that the tomographic analysis can significantly improve constraints on cosmological and IA parameters.  \citep{king2003separating, heymans2004weak, troxel2015intrinsic}. Although it is straightforward to generalize all the results shown in this paper to the tomographic case, this will require increasing the resolution of the simulation at lower redshifts (as illustrated by \autoref{fig:clsktng_comp}) in order to model correctly low redshift bins. Since the maximum number of particles we can adopt in a simulation is closely limited to the GPU memory, we are building a distributed implementation of DLL, which will allow us to increase the resolutions of the simulations to the point of modeling correctly even the smaller scales at the lower redshifts.

Another direction for further development is the ray tracing methodology. In our method, we construct the weak lensing maps assuming the Born approximation. However, \cite{petri2017validity} shows that for an LSST-like survey, while the Born approximation leads to negligible parameter bias for the power spectrum, it can lead to significant parameter bias for higher-order statistics. 
Hence, the natural next step will be to implement a ray-tracing algorithm beyond the Born approximation in our pipeline.
 We aim to adopt the multiple-lens-plane approximation \citep{1986ApJ...310..568B, Stella_Seitz_1994, jain2000ray, Vale_2003, refId0}, i.e. by introducing lens planes perpendicular to the line-of-sight, the deflection experienced by the light rays due to the matter inhomogeneities will be approximated through multiple deflections at the lens planes. 
More specifically, we will implement the memory-efficient ray-tracing scheme proposed by \cite{refId0} in the Tensorflow framework.

On the theoretical modeling side, we studied the impact of the intrinsic alignment of galaxies assuming a linear coupling between the intrinsic galaxy shapes and the non-linear projected tidal fields, i.e. adopting the NLA model. This physical description for the IA is only an approximation since it does not take into account the tidal torque field.
In future work, we aim to extend the NLA model by implementing the extended $\delta$NLA model, described by \cite{harnois2021cosmic}.

Finally, we presented a tool based on only Dark matter simulations. We note that this would force us to perform conservative scale cuts in the inference analysis to not include scales affected by baryonic effects.
 A future prospect is to include baryonic effects in the analysis. One possible way applicable to our methodology  could be to extend the Hybrid Physical Neural ODE approach and apply more sophisticated models to learn the physics that controls the hydrodynamics simulations.

We expect that the methods illustrated in this paper will be extended to different relevant use-cases. A particularly suitable example is related to the application of algorithms such as the Variational Inference and the Hamiltonian Monte Carlo that are widely used in the Bayesian inference context and were until now excluded due to the lack of derivatives. 
A further example is provided by \cite{zeghal2022neural}, which demonstrates that having access to the gradients of the forward model is beneficial to constrain the posterior density estimates.

\section{Conclusions}\label{Conclusions}
In this paper, we have presented the Differentiable Lensing Lightcone (DLL) model, a fast lensing lightcone simulator providing access to the gradient. 
We extended the public FlowPM N-body code, implementing the Born approximation in the Tensorflow framework to create non-Gaussian convergence maps of weak gravitational lensing. To allow DLL to run at low resolution without affecting significantly the accuracy, we complement the FlowPM N-body code with the Hybrid-Physical Neural scheme, a new correction scheme for quasi N-body PM solver, based on Neural Network implemented as a Fourier-space filter. 
We validate our tool by comparing the $C_{\ell}$ and peak counts statistics against predictions from $\kappa$TNG simulations. 
To do this, we run simulations following the evolution of $128^3$ particles and we produce weak lensing convergence maps for several redshift sources. 
We show that, despite being generated at low computation costs, we recover a good match for redshift equal or higher than $z=0.91$. 
To demonstrate the potential of our tool, as a first use case, we exploit the automatic differentiability of the simulations to do Fisher forecast. 
Thanks to back-propagation, accessing the derivative through the simulations w.r.t. the cosmological parameters and $A_{IA}$ parameter is possible at the same computational cost of the forward simulation.
Assuming an LSST-like setting, we simulate weak lensing convergence maps for a single source redshift $z=0.91$ and angular extend of 5$^\circ$, based on a periodic box of
comoving volume equal to  $205$ $h^{-1}$Mpc.
We compute the constraints on the resulting convergence maps
with the starlet peak counts and use a wavelet-filtered lensing power spectrum as a benchmark for the comparison.
 Within the limits of the analysis choices made in this study, we obtain the following results:
\begin{itemize}
    \item  We confirm that the peak count statistics outperform the two-point statistics as found in \cite{ajani2020constraining},
 even in high dimensional cosmological and nuisance parameter space.
    \item We find the peak counts to provide the most stringent constraints on the galaxy intrinsic alignment amplitude $A_{IA}$. 
\end{itemize}

To conclude, the framework outlined here can provide many advantages in the context of cosmological parameter inference:  it is the first step in the development of fully differentiable inference pipelines for weak lensing, it is a fast tool to further explore the sensitivity of higher-order statistics to systematics. 

\begin{acknowledgements}

DESC acknowledges ongoing support from the IN2P3 (France), the STFC 
(United Kingdom), and the DOE, NSF, and LSST Corporation (United States).  
DESC uses resources of the IN2P3 Computing Center 
(CC-IN2P3--Lyon/Villeurbanne - France) funded by the Centre National de la
Recherche Scientifique; the National Energy Research Scientific Computing
Center, a DOE Office of Science User Facility supported under Contract 
No.\ DE-AC02-05CH11231; STFC DiRAC HPC Facilities, funded by UK BEIS National 
E-infrastructure capital grants; and the UK particle physics grid, supported
by the GridPP Collaboration.  This work was performed in part under DOE 
Contract DE-AC02-76SF00515.
This work was granted access to the HPC/AI resources of IDRIS under the allocation 2022-AD011013922 made by GENCI.
BH is supported by the AI Accelerator program of the Schmidt Futures Foundation; JHD acknowledges support from an STFC Ernest Rutherford Fellowship (project reference ST/S004858/1); The presented work used computing resources provided by DESC at the National Energy Research Scientific Computing Center, a DOE Office of Science User Facility supported by the Office of Science of the U.S. Department of Energy under Contract No. DEAC02-05CH11231; 
This paper has undergone internal review in the LSST Dark Energy Science Collaboration. We thank the internal reviewers, Virginia Ajani, William Coulton, and Jia Liu for their constructive comments.
Author contributions to this work are as follows: Denise Lanzieri led the analysis and the writing of the paper. Fran\c{c}ois Lanusse designed the project, contributed to code and text development.
Chirag Modi contributed to the PM code and text development.
Benjamin Horowitz contributed to the text and made minor code developments. Joachim Harnois-Déraps contributed to the intrinsic alignment modeling, provided general advice on simulations, and contributed to the text.
Jean-Luc Starck provided expertise on wavelet peak counting and higher-order weak lensing statistics. 
\end{acknowledgements}

%
%
\bibliographystyle{aa} 
\bibliography{biblio} 
\begin{appendix}

\section{Validation against a theory prediction}
We show the $2\sigma$ constraints obtained from our Fisher analysis of standard $C_{\ell}$ (orange) and the theoretical prediction from halofit (blue contours) in \autoref{fig:Fisher_result_theory}. The dashed black contours define the prior used for the forecasting.
The analysis is performed for one single source redshift at z=0.91 and the survey setup presented in \autoref{Validating_simulations_for_LSST_Y}.

The constraints from the theoretical predictions
are compatible with the ones obtained from the mock DLL maps for all cosmological parameters, except $n_s$. Indeed, despite sharing the same direction of the degeneracy, the theoretical contours for $n_s$ are narrower compared to the ones obtained in our analysis. In general, this translates into an increased uncertainty in constraining $n_s$, most likely due to the deficit in power observed for the $C_{\ell}$ at small scales. 

The theoretical predictions are computed using the public library \texttt{jax-cosmo} \citep{Campagne_2023}. 
We want to highlight that both the theoretical Fisher matrices and the ones from our analysis are obtained by automatic differentiation.

\renewcommand{\thefigure}{A\arabic{figure}}
\setcounter{figure}{0}
 \begin{figure*}
    \centering
    \includegraphics[width=\textwidth]{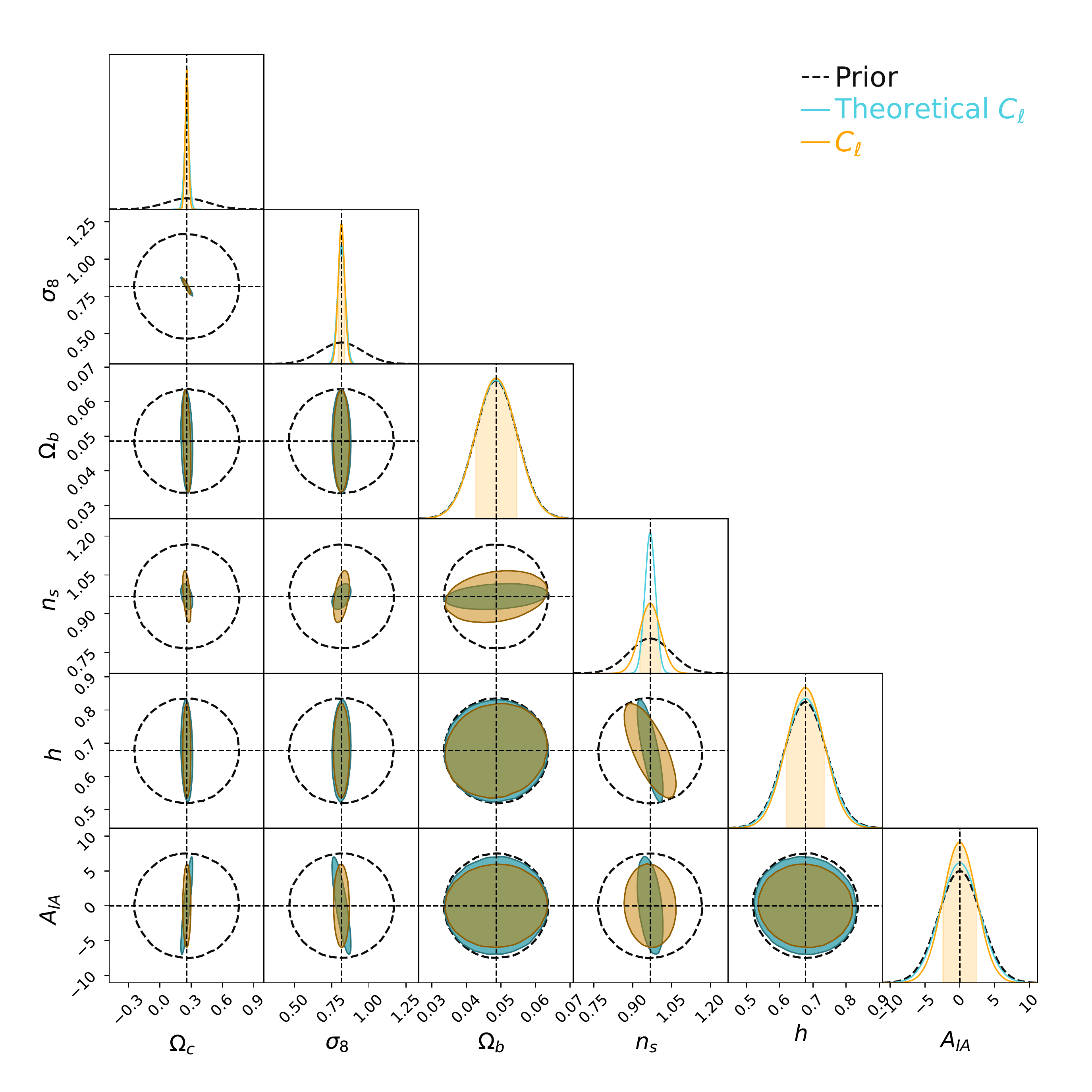}
    \caption{ $2\sigma$ contours derived for one single source redshift at z=0.91 and the survey setup presented in \autoref{Validating_simulations_for_LSST_Y}.
    We compare the Fisher matrix constraints on cosmological parameters and $A_{IA}$ amplitude obtained with the $C_{\ell}$ from our mock maps (orange) and the theoretical $C_{\ell}$ (blue) obtained from the public library  \href{https://github.com/DifferentiableUniverseInitiative/jax_cosmo}{\texttt{jax-cosmo}} \citep{Campagne_2023}.  
    In both cases, the constraints are obtained by applying  the wavelet pass-band filter for the $C_{\ell}$ as described in \autoref{Validating_simulations_for_LSST_Y}.
The dashed black contours are the prior used for the forecasting.}
     \label{fig:Fisher_result_theory}
\end{figure*}

\section{Validate the stability of the Fisher contours}
To ensure that the shape of the ellipses and the direction of the degeneracies are not the results of stochasticity, we prove the stability of the Fisher analysis results by testing the convergence of the jacobians.
In \autoref{fig:Fisher_stab} we present the results for Fisher constraints obtained when varying the number of independent simulations used to compute the jacobians.  As we can see, the convergence seems to be reached for the peak counts with, at least, 1500 realizations. On the other hand, the angular $C_{\ell}$ proves to be more sensitive to noise, and thus requires at least 2600 realizations.

\renewcommand{\thefigure}{B\arabic{figure}}
\setcounter{figure}{0}
 \begin{figure*}
    \centering
        \includegraphics[width=0.65\textwidth]{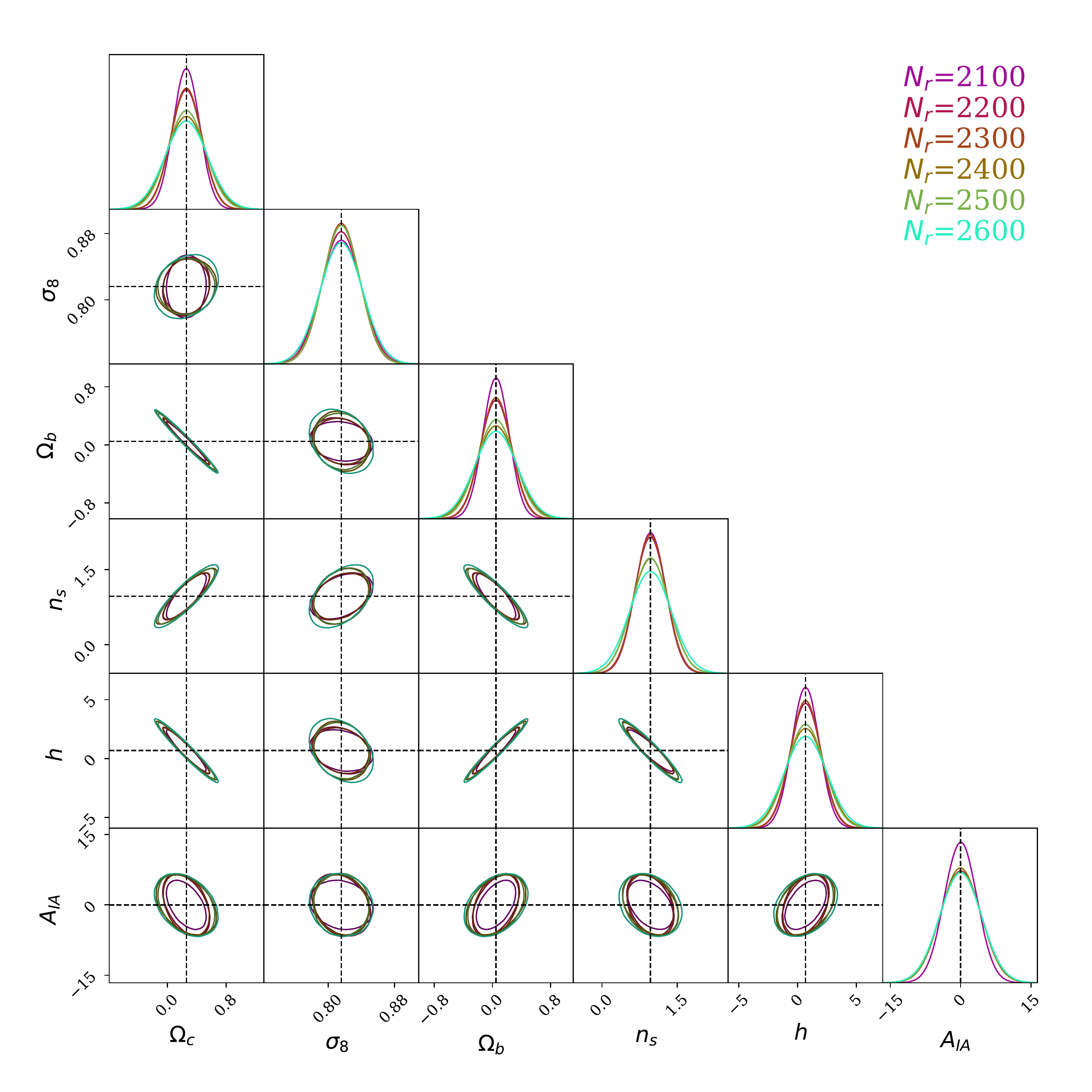}
        \includegraphics[width=0.65\textwidth]{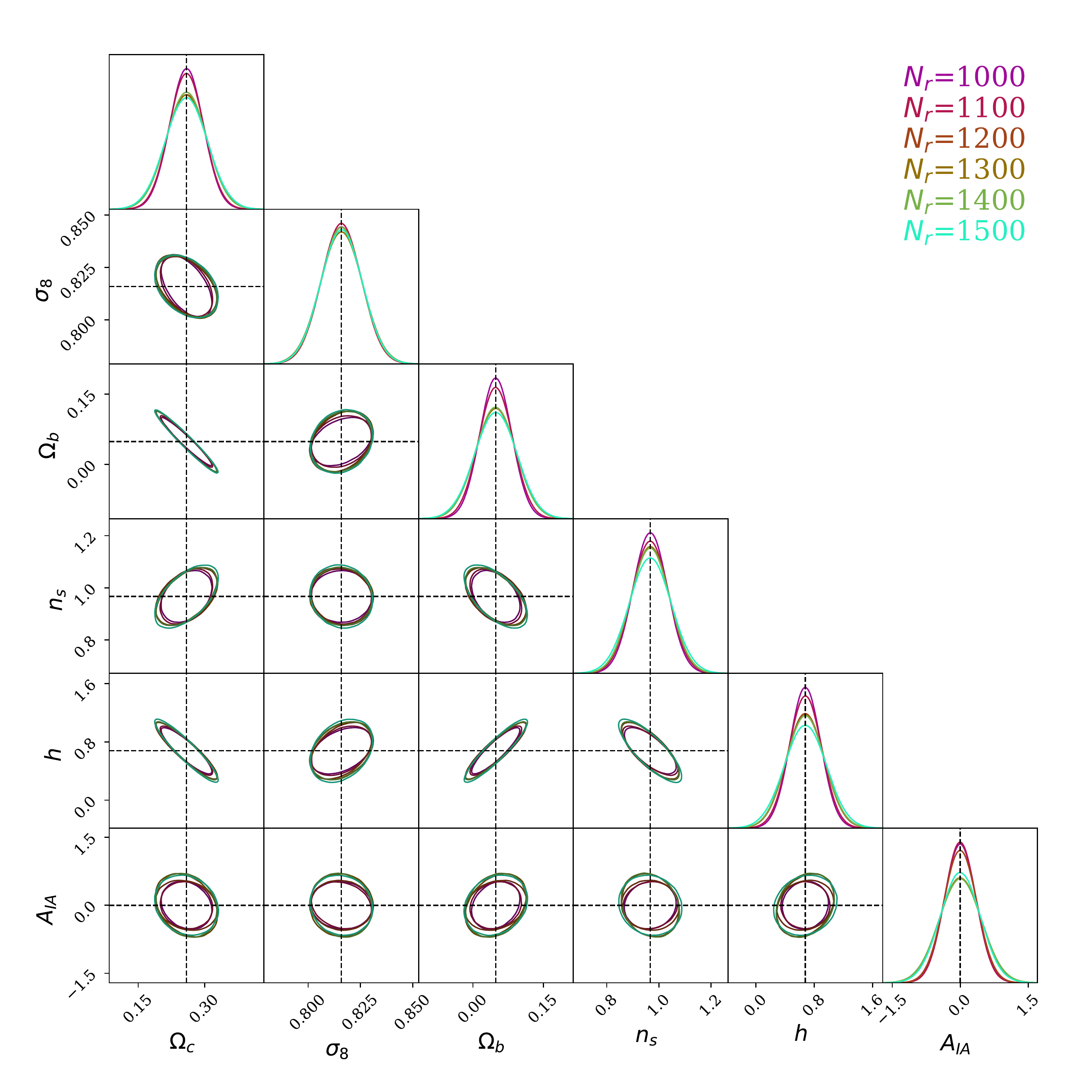}
    \caption{ $1\sigma$ Fisher contours derived for one single source redshift at z=0.91 and the survey setup presented in \autoref{Validating_simulations_for_LSST_Y} for the $C_{\ell}$ (upper panel) and the Peak counts (lower panel). The different colors refer to the number of independent realizations used to mean the Jacobian in the \autoref{Fisher_matrix}. The dashed black lines are located at the fiducial parameter values.}
     \label{fig:Fisher_stab}
\end{figure*}

\section{Validation with higher resolutions simulations}
In \autoref{Validating_simulations_for_LSST_Y}, we presented a validation of our simulations by comparing the statistics from DLL and $\kappa$TNG-Dark.
In particular, we have seen a discrepancy on small scales for both the $C_{\ell}$ and the peak counts. We have attributed this bias to the low resolution of the simulations. Therefore, to justify this assumption, we simulate convergence maps of higher resolution, i.e, we raise the number of particles but keep the same box of 205$^3$($h^{-1}$Mpc$)^3$ in the simulation. In the left panel of  \autoref{fig:ps_comp_high}, we present the angular power spectrum computed from our DLL with the original number count of particles ($128^3$) and the power spectrum computed from higher resolutions DLL simulations ($212^3$). The two outputs are compared to the $\kappa$TNG predictions. In the right panel  of \autoref{fig:ps_comp_high} the fractional differences between the convergence power spectra from the two maps and the $\kappa$TNG are shown. We can see that, by increasing the number of particles, we can improve the accuracy of the lensing $C_{\ell}$ up to $~20\%$.
In \autoref{peaks_high}, we compare the peak counts statistic from our map to the one from the $\kappa$TNG for different resolutions. We use the same wavelet decomposition presented in \autoref{peak_sec}. As for the power spectrum, we note the same tendency to recover better accuracy when the resolution is increased. 

Finally, we reproduce the results of the Fisher analysis for the intrinsic alignment parameter $A_{IA}$ with higher-resolution simulations. We adopt the same Forecast criteria presented in \autoref{Application_Fisher_forecast_for_HOS}. However, for this specific test, we compute the derivatives numerically using the finite differences. The step sizes used for these variations are $\Delta x_{IA}=0.15$ for the $C_{\ell}$ and $\Delta x_{IA}=1.2$ for the peak counts. 
In order to check the reliability of the numerical derivatives, we investigate the stability of the Fisher forecast against different step sizes used to compute them.
The derivatives are computed as the mean of 3000 independent realizations for both $C_{\ell}$ and peak counts. 

 We confirm the peak counts provide the most stringent constraints on the galaxy intrinsic alignment amplitude $A_{IA}$.
As for the full analysis, we tested the stability of the Fisher forecast by varying the number of simulated maps used to compute the derivatives. In \autoref{tuning_high} we present the $1\sigma$ error on $A_{IA}$ when varying the number of independent realizations used to compute the derivatives. It is interesting to note, that, even in this case, the derivatives of the $C_{\ell}$ can not be considered fully converged. However, as can be noted from the stability plots of \autoref{tuning_high}, the noise in the derivatives leads to tighter constraints in the Fisher forecast. Hence, the fully converged derivatives of the $C_{\ell}$ would result in even broader constraints, without changing the results we found.

\renewcommand{\thefigure}{C\arabic{figure}}
\setcounter{figure}{0}
\begin{figure*}
    \centering
    \includegraphics[width=\columnwidth]{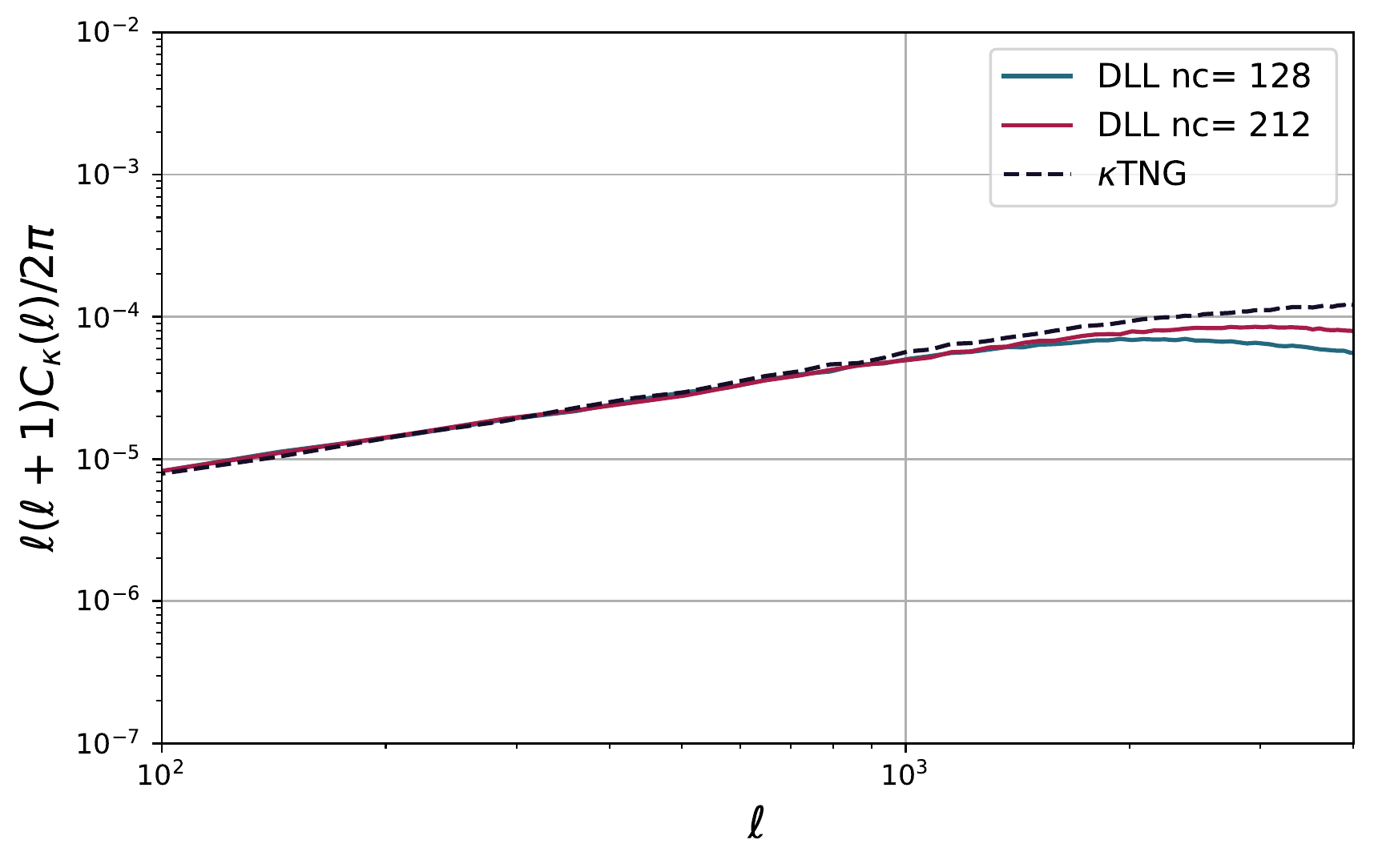}
        \includegraphics[width=\columnwidth]{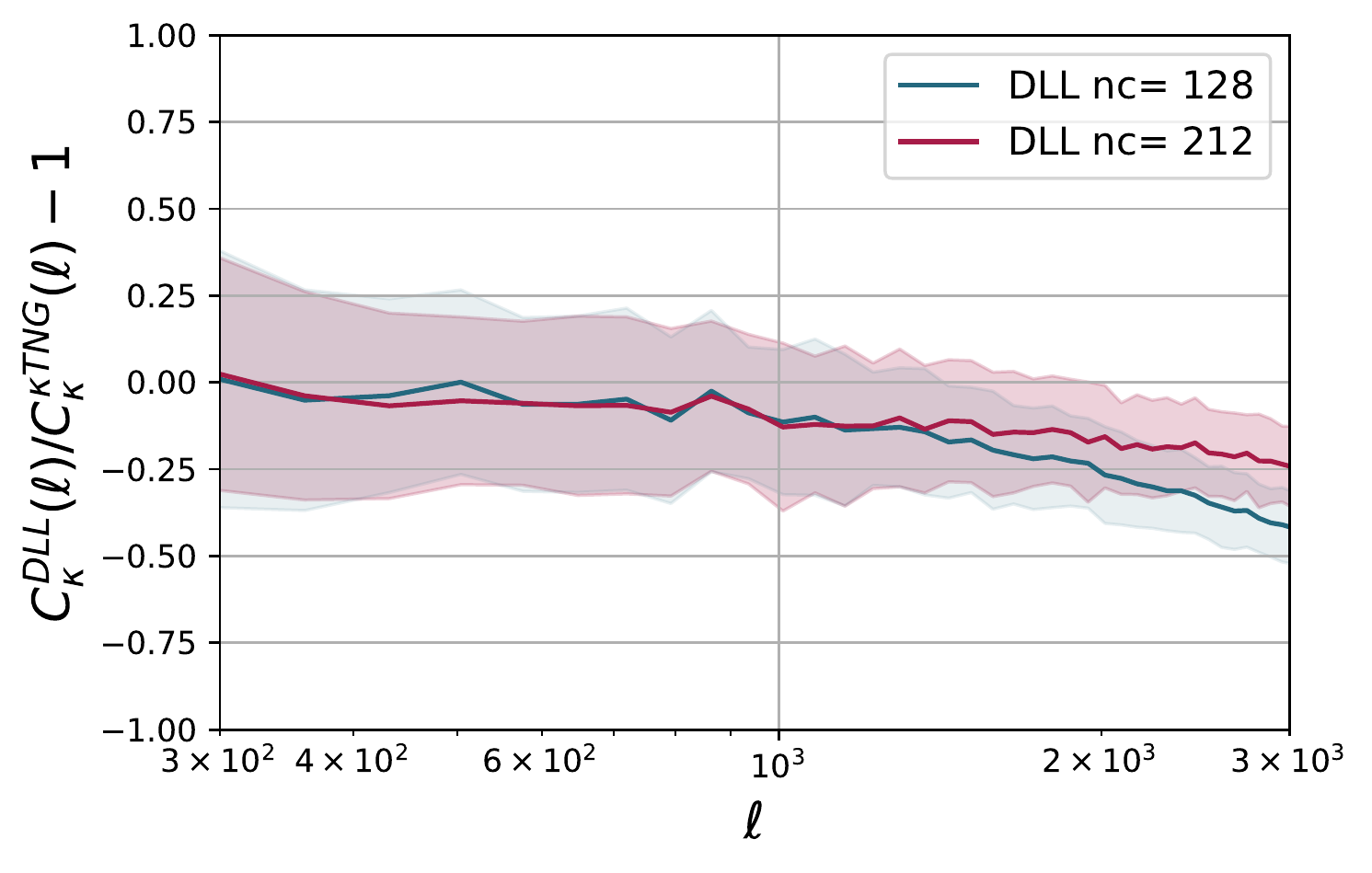}
    \caption{
    \textbf{Left panel}: Angular power spectra of PM simulations with original resolution (number count 128$^3$, blue line) and higher resolution (number count 212$^3$, red line) compared to the $\kappa$TNG prediction. \textbf{Right panel}: fractional angular power spectrum of PM simulations with original and higher resolution and the $\kappa$TNG prediction. The power spectra and ratios are means over 100 independent map realisations and the shaded regions represent the error on the mean. The spectra are computed for the source redshift $z_s=0.91$. 
}
    \label{fig:ps_comp_high}
\end{figure*}

\renewcommand{\thefigure}{C\arabic{figure}}
\setcounter{figure}{1}
\begin{figure*}\label{peaks_high}
    \centering
    \includegraphics[width=\columnwidth]{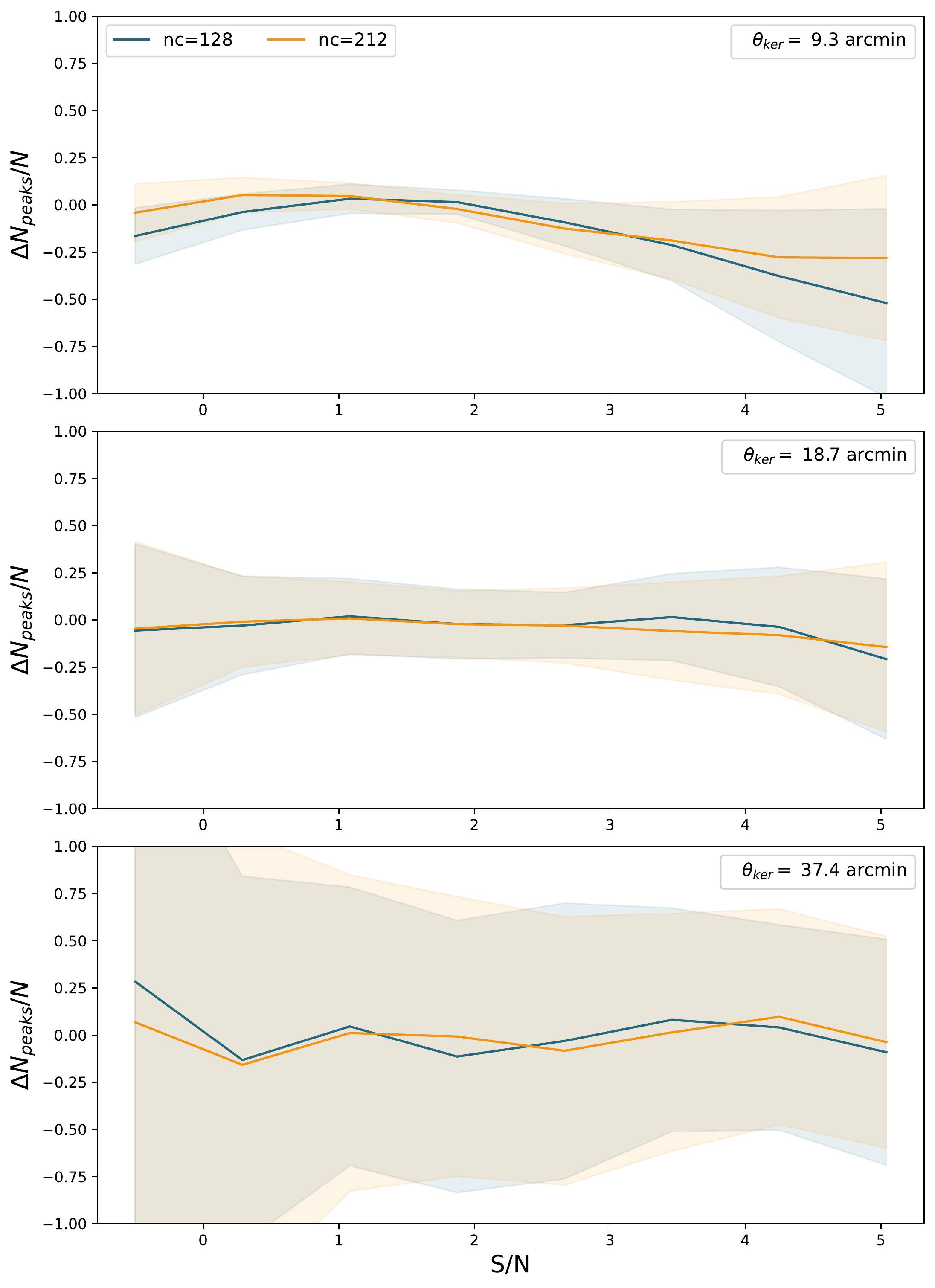}
    \caption{Fractional number of peaks of DLL simulations and $\kappa$TNG simulations. The results are shown for the number counts $128^3$ (blue lines) and $212^3$ (orange lines). The peak counts distributions are shown for each starlet scales resolutions used: 9.34 (upper panel), 18.17 (center panel), 37.38 arcmins (lower panel).
  The results mean over 100 independent map realisations, the shaded regions represent the error on the mean. The statistics are computed for the source redshift $z_s=0.91$. }
\end{figure*}

\renewcommand{\thefigure}{C\arabic{figure}}
\setcounter{figure}{2}
\begin{figure*}\label{tuning_high}
    \centering
     \includegraphics[width=\columnwidth]{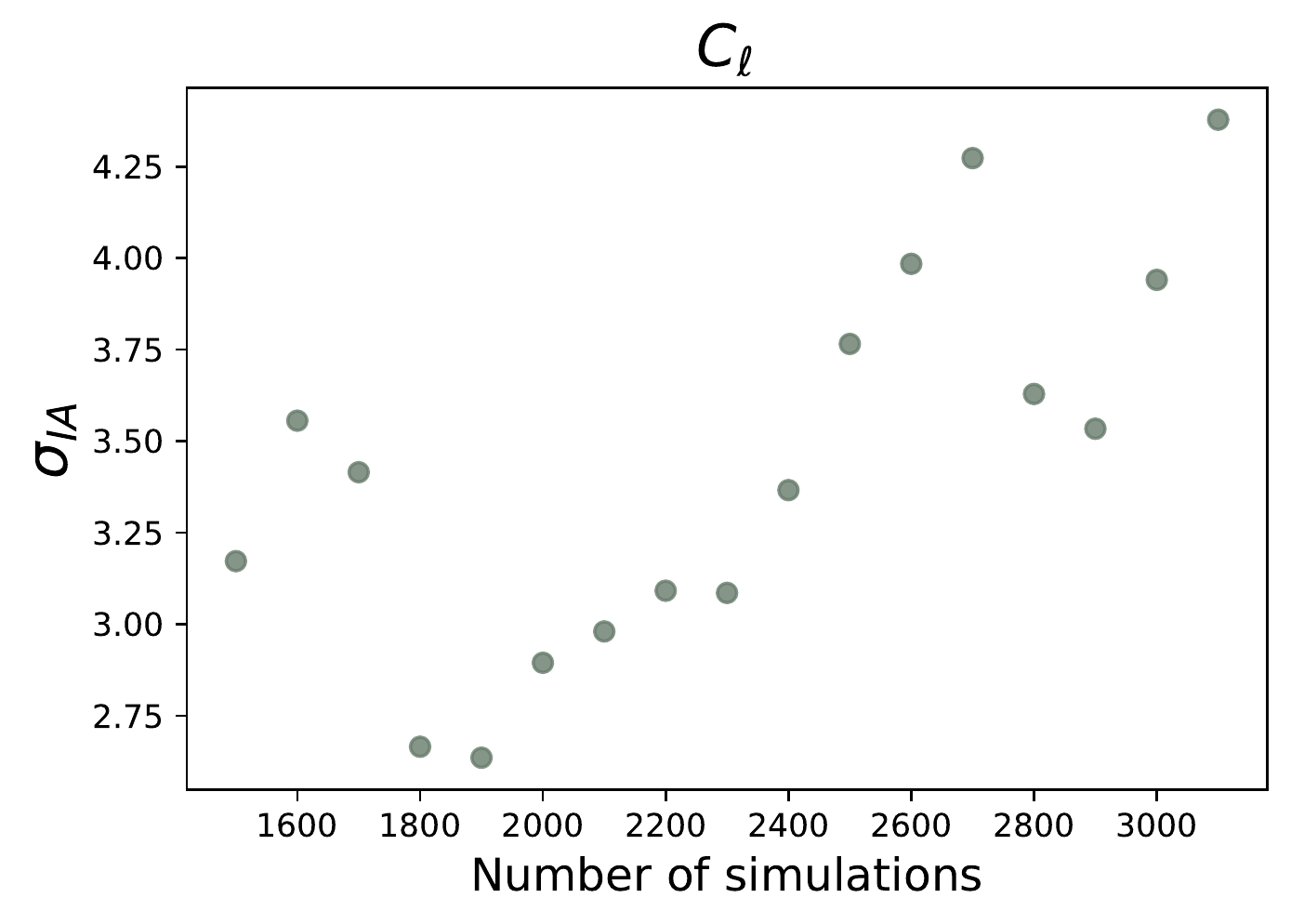}
        \includegraphics[width=\columnwidth]{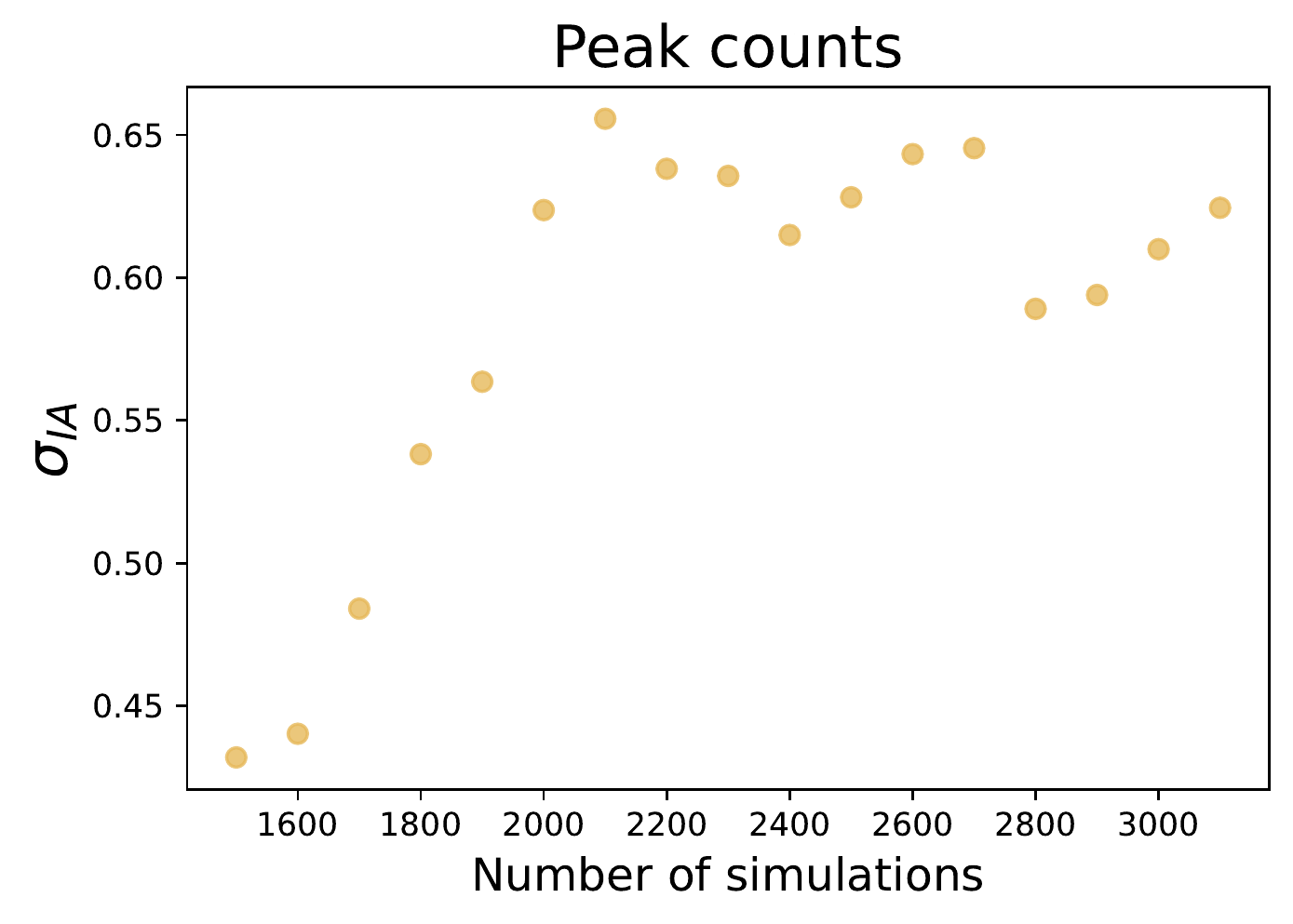}
    \caption{
    $1\sigma$ error on $A_{IA}$ derived for one single source redshift at z=0.91 for different numbers of independent realizations used to mean the Jacobian in the \autoref{Fisher_matrix}. The results are shown for the $C_{\ell}$ (left panel) and the peak counts (right panel).  
}
\end{figure*}

\renewcommand{\thefigure}{C\arabic{figure}}
\setcounter{figure}{3}
\begin{figure*}
    \centering
    \includegraphics[width=\textwidth]{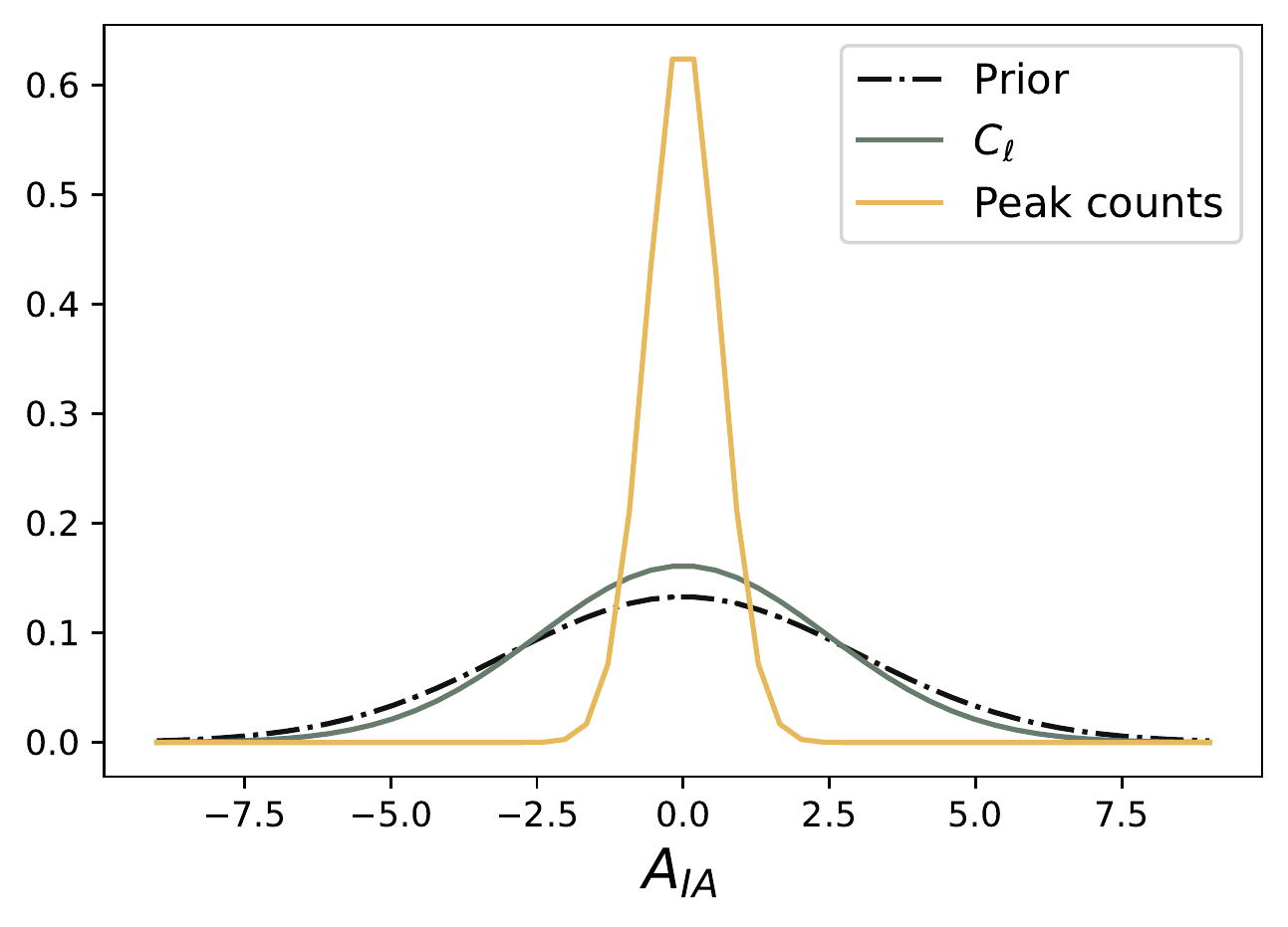}
    \caption{$1\sigma$ error on $A_{IA}$ for one single source redshift at z=0.91 and the survey setup presented in \autoref{Validating_simulations_for_LSST_Y}, from mock simulations with $212^3$ particles. The results are obtained by applying the starlet Peak counts (yellow contours) computed on noisy maps filtered using a starlet kernel of [$9.33, 18.70, 37.38$] arcmin together and the wavelet pass-band filter for the $C_{\ell}$ statistics (grey contours) as described in \autoref{Validating_simulations_for_LSST_Y}.  }
     \label{fig:Fisher_result_ia}
\end{figure*}

\end{appendix}
\end{document}